\documentclass[aps,showpacs]{revtex4}
\usepackage[utf8]{inputenc}
\usepackage{color}
\usepackage{amsmath}
\usepackage{graphicx}

\usepackage{hyperref}

\makeatletter
\@ifundefined{textcolor}{}
{%
 \definecolor{BLACK}{gray}{0}
 \definecolor{WHITE}{gray}{1}
 \definecolor{RED}{rgb}{1,0,0}
 \definecolor{GREEN}{rgb}{0,1,0}
 \definecolor{BLUE}{rgb}{0,0,1}
 \definecolor{CYAN}{cmyk}{1,0,0,0}
 \definecolor{MAGENTA}{cmyk}{0,1,0,0}
 \definecolor{YELLOW}{cmyk}{0,0,1,0}
}

\usepackage{bm}
\makeatother

\begin{document}
\title{Quantum coherence, quantum Fisher information and teleportation in the Ising-Heisenberg spin chain model of a heterotrimetallic $\mathrm{Fe-Mn-Cu}$ coordination polymer with magnetic impurity}
\author{Hamid Arian Zad$^{1,2}$, Moises Rojas$^{3}$}
\affiliation{$^{1}$A.I. Alikhanyan National Science Laboratory, 0036, Yerevan, Armenia}
\affiliation{$^{2}$ICTP, Strada Costiera 11, I-34151 Trieste, Italy}
\affiliation{$^{3}$Departamento de Física, Universidade Federal de Lavras, 37200-900,
Lavras-MG, Brazil}

\begin{abstract}
The effect of non-uniform magnetic fields on the Ising-Heisenberg chain of a heterotrimetallic coordination compound $\mathrm{Fe-Mn-Cu}$,
modeling a magnetic impurity on one dimer is studied. This impurity is configured by imposing non-uniform magnetic fields on each sites of
 $j-$th interstitial ionic dimer $\mathrm{Mn}^{2+}-\mathrm{Cu}^{2+}$ of the chain model.
The quantum coherence and pairwise entanglement between spin-1/2 magnetic $\mathrm{Mn}^{2+}-\mathrm{Cu}^{2+}$  ion dimers clearly depend on the site which the impurity is located. It is demonstrated that when the magnetic impurity is considered for one magnetic dimer, by altering Ising nodal exchange interaction and Heisenberg anisotropy parameter of the impurity dimer, the entanglement can be enhanced to the maximal value 1 at a special fixed magnetic field. Moreover, we find that the quantum Fisher information of the model with magnetic impurity behaves considerably different from the original model. Besides of the quantum resources like concurrence, we prove that the quantum Fisher information can be used as new quantum tool for estimating the quantum phase transition in the model under consideration.
On the other hand, thermal teleportation can be significantly optimized by adjusting the magnetic impurity, and a strong increase in the average fidelity is observed.  Finally, the magnetic impurity can be manipulated  to  locally control the thermal entanglement, coherence, quantum Fisher information and teleportation unlike the original model where it is done globally.
\end{abstract}
\maketitle

\section{Introduction}

In quantum resource theories \cite{bru,ade,chi,Guff2019}, the exploration
of quantum correlations plays an important role in different fundamental
aspects in quantum information processing. The quantum coherence is
based on quantum superposition and closely connected to quantum correlations,
which is the origin of many quantum phenomena such as the interference
of light \cite{Ficek2005,Walls2008,Cimini2019}, lasers \cite{Serapiglia2000}, superconducting Josephson junctions \cite{Cleland2004,Bauch2005,Ferrini2008,You2011}, and superfluidity \cite{Volovik2008,Narayana2011,Anankine2017}.
 Further applications include quantum thermodynamics \cite{bran,va}, quantum biology \cite{huel,lam} 
and quantum metrology \cite{mac,fro}. 
 Several intensive attempts have been made to investigate the interrelations between
coherence and entanglement \cite{tan,Chitambar2016,mondal,Kim2018}, as well as, between coherence and discord
\cite{yao,xi,Hu}. As a beneficial recourse, in Ref. \cite{Streltsov2015},  A. Streltsov {\it et al.} have quantitatively and operationally provided a quite clear connection between coherence and entanglement that generally allows one to formulate a novel method to quantify coherence in terms of entanglement. 

 In recent years, various measures of coherence have been proposed, and their properties have been investigated in detail.
 For example, in Ref. \cite{baum}, the authors introduced the $l_1-$norm of coherence and the relative entropy of coherence as measures of coherence and proved that every proper measure of coherence should satisfy the  $l_1-$norm of coherence and relative entropy of coherence 
  as the most general and easy-to-use coherence quantifiers.
  Next, P. Zanardi {\it et al.} introduced a coherence matrix by which one can define a wide variety families of measures of the
coherence \cite{Zanardi2017}.   
In the present work, we confine ourselves to utilize the $l_1-$norm of coherence.

The Heisenberg spin chain is one of the simplest quantum systems which
exhibits entanglement. Due to this fact, in the current decade, the
thermal entanglement and quantum teleportation in various Heisenberg spin models has been rigorously studied
\cite{zhang,bow,Arian2017}.
Motivated by real material $\mathrm{Cu}_{3}\left(\mathrm{CO}_{3}\right)_{2}\left(\mathrm{OH}\right)_{2}$
known as \textit{azurite} \cite{kiku},  which can be properly characterized
by Heisenberg model on a generalized diamond chain, 
several Ising-Heisenberg diamond chains have been examined
\cite{joz,joz-1,rojas}. In particular, the thermal entanglement
properties were extensively investigated \cite{moi,cheng,moi-1,xu,moi-2}
and, more recently, the quantum teleportation through a couple of
Ising-XXZ diamond chain \cite{moises} have also been reported. 

On the other hand, motivated by the heterotrimetallic coordination compound 
$[\mathrm{Cu}^{\mathrm{II}}\mathrm{Mn}^{\mathrm{II}}(\mathrm{L}^1)][\mathrm{Fe}^{\mathrm{III}}(\mathrm{bpb})(\mathrm{CN})_2]\cdot\mathrm{ClO}_4\cdot\mathrm{H}_2\mathrm{O}$, abbreviated by simple notation 
$\mathrm{Fe-Mn-Cu}$ \cite{wang},
 the magnetization process and thermal entanglement of such a polymer on the 1-D Ising-Heisenebrg spin model have already been investigated in detail \cite{souza}.
Not long ago, the entanglement teleportation via a couple of quantum
channel based on the Ising-Heisenberg spin configuration of heterotrimetallic $\mathrm{Fe-Mn-Cu}$
coordination polymer was examined so far \cite{zheng}.  A very similar study has been carried out on the heterobimetalic coordination polymer
$[(\mathrm{Tp})_{2}\mathrm{Fe}_{2}(\mathrm{CN})_6(\mathrm{OAc})(\mathrm{bap})\mathrm{Cu}_2(\mathrm{CH}_3\mathrm{OH})\cdot2\mathrm{CH}_3\mathrm{OH}\cdot\mathrm{H}_2\mathrm{O}]$ in Ref. \cite{Jozef2019}, where the compound has different spin configuration.

The quantum Fisher information (QFI) is introduced as the quantum version of the Cram{\' e}r-Rao inequality \cite{Braunstein}. It provides the possibility of estimating the quantum metrology, which quantifies the maximum precision, namely,  Cram{\' e}r-Rao bound makes a tight connection between QFI and estimation theory \cite{Bakmou2019,Slaoui2019,Jafari2020}. With a given quantum state $\rho$, one can estimate QFI of a special spin model by manipulating the eigenvalues and eigenvectors of the density matrix $\rho$. 
In some recent studies it has been demonstrated that besides of quantum metrology, the QFI can be admittedly connected to the quantum phase transition \cite{Marzolino2017,Liu2020}. For instance, the quantum phase transition of the XXZ Heisenberg model has been studied through QFI by Biao-Liang Ye {\it et al.} in Ref. \cite{Ye2020}.
 Not so long ago, the same author with contributing other co-workers, in Ref. \cite{Ye2018}, have investigated the quantum phase transition in the XY spin models possessing Dzyaloshinsky-Moriya interaction via QFI.
Here, we plan to verify the QFI for the Ising-Heisenberg spin chain model of the heterotrimetallic $\mathrm{Fe-Mn-Cu}$ coordination polymer with magnetic impurity and look for critical point(s) at which quantum phase transition occurs. Our work is original and is one of the first-time researches devoted to the subject of  spin models with impurities.

Impurities play essential  role in the quantum spin models, such that a tiny defect may destroy the properties of the system \cite{Fu2002,Sun2017}. It  is then pragmatic full to investigate the entanglement of spin models possessing impurity. 
%In the previous studies, impurity effects on quantum entanglement have been considered for the diamond chains so far  \citep{mo-1,mo-2}. 
%However, the entanglement in Heisenberg spin chain with multi-impurity on the period boundary condition has been less studied.
There has been considerable activity in the study of spin structures with spin impurities in the literature. In particular,
the spin chains with impurities can be viewed as  models with
site imperfection or impurity. The strength of interactions between
the impurity spin and its neighboring spins can be different from
that between the normal spins \cite{plas,fal,fuku}. The
thermal quantum correlations of the Ising-Heisenberg model on diamond-like
chain with magnetic impurities have been discussed in  earlier notes \cite{mo-1,mo-2}. 
%Besides of spin impurity we have also considered the effect of the nonuniform magnetic field -magnetic impurity- in such diamond chain models. 
These models
describes the situation in which a magnetic impurity is located on
one or more spins of the spin chain structure. The impurity effects
on quantum entanglement, discord \cite{tony,hon,galin,sha,tie,zv,ji} and NMR \cite{zv2} have been considerably investigated in Heisenberg spin chains. 
%Moreover,  in Heisenberg chain with magnetic impurity has already been discussed \citep{ji}.

In the present paper, our main goal is to investigate the thermal
entanglement, quantum coherence, and quantum teleportation in an exactly
solvable spin-1/2 Ising-Heisenberg of the heterotrimetallic coordination
compound $\mathrm{Fe-Mn-Cu}$ with magnetic impurity. Furthermore, we will examine the QFI and its first magnetic field derivative for the model under consideration with magnetic impurity.
In this approach, the magnetic impurity is located on the one plaquette of the spin chain
and can serve as an efficient way of controlling the quantum correlations.
 Controlling and modifying  the quantum correlations is carried out locally upon modulating the magnetic field. 
 In addition, we will explicitly show that a superb enhancement in teleportation can be achieved by tuning the strength of the magnetic impurity.

This paper is organized as follows. In Section \ref{Model}, we describe the
physical model and the method of its exact treatment. In Section \ref{QC},
is given a brief review concerning the definition of concurrence,
quantum coherence and QFI. Furthermore an analytical expression is found for the concurrence
and quantum coherence. Then, we discuss the thermal
entanglement, quantum coherence and QFI of Heisenberg dimer with magnetic
impurity, in this section. In Section \ref{QT}, we discuss the effects of the magnetic impurity
in the quantum teleportation. Finally, we conclude in Section \ref{Conclusions}.  

\section{The model and method}\label{Model}
\begin{figure}
\includegraphics[scale=0.7,trim=10 160 10 100, clip]{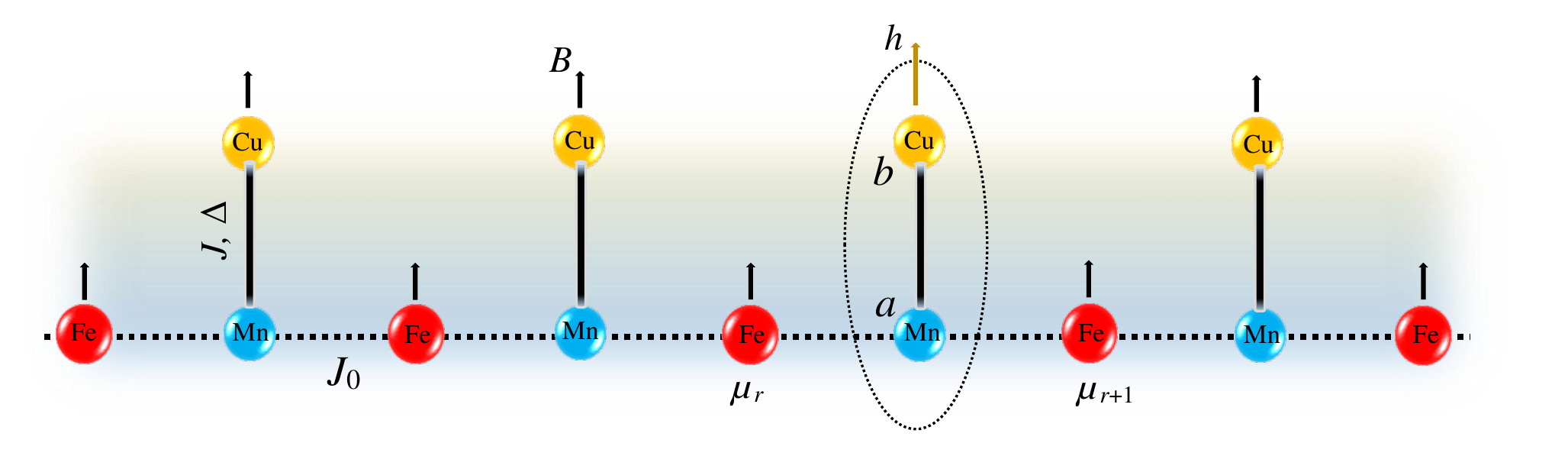}\caption{\label{fig:model}
A schematic representation of the  Ising-Heisenberg spin chain such that the exchange interaction between 
$\mathrm{Fe}^{3+}-\mathrm{Mn}^{2+}$ ions is assumed to be Ising-type interaction, whereas each pair of
 $\mathrm{Mn}^{2+}-\mathrm{Cu}^{2+}$ magnetic ions interact together via an anisotropic Heisenberg exchange interaction. The magnetic impurity is rounded by elliptic dotted line. }

\end{figure}

In this section, let us consider the heterotrimetallic coordination compound $[\mathrm{Cu}^{\mathrm{II}}\mathrm{Mn}^{\mathrm{II}}(\mathrm{L}^1)][\mathrm{Fe}^{\mathrm{III}}(bpb)(\mathrm{CN})_2]\cdot\mathrm{ClO}_4\cdot\mathrm{H}_2\mathrm{O}$ ( abbreviated by simple notation 
$\mathrm{Fe-Mn-Cu}$) as a magnetic system on a one-dimensional spin chain schematically represented in Fig. \ref{fig:model}.
 The total Hamiltonian of the model containing a magnetic impurity can be given by $\mathcal{H}=\sum_{i=1}^{N}\mathcal{H}_{i}$, where

\begin{equation}
\mathcal{H}_i=\mathcal{H}_{i}^{host}+\mathcal{H}_{i}^{imp},\label{eq:1}
\end{equation}
The host Hamiltonian $\mathcal{H}_{i}^{host}$ reads

\[
\begin{array}{cl}
\mathcal{H}_{i}^{host}= J\left(\mathbf{S}_{a,i},\mathbf{S}_{b,i}\right)_{\Delta}+J_{0}S_{a,i}^{z}\left(\mu_{i}+\mu_{i+1}\right)
  -B_{2}S_{a,i}^{z}-B_{3}S_{b,i}^{z}-\frac{B_{1}}{2}\left(\mu_{i}+\mu_{i+1}\right),\\
  \\
  \mathrm{for}\:i=1,2,\ldots,r-1,r+1,\ldots,N.
\end{array}
\]
Above, $S_{a,i}^{\alpha}$ and $S_{b,i}^{\alpha}$ with $\alpha=\{x,y,z\}$ signify the Heisenberg spin$-\frac{1}{2}$ operators of, respectively, 
$\mathrm{Mn}^{2+}$ and $\mathrm{Cu}^{2+}$ magnetic ions. 
The quantum part $({\boldsymbol S}_{a,i}\cdot{\boldsymbol S}_{b,i})_{\Delta}$ that corresponds to the interstitial Heisenberg dimer interactions $J$ and $\Delta$ can be defined by  
\begin{equation}
\begin{array}{lcl}
 ({\boldsymbol S}_{a,i}\cdot{\boldsymbol S}_{b,i})_{\Delta}\equiv S_{a,i}^xS_{b,i}^x+S_{a,i}^yS_{b,i}^y+\Delta S_{a,i}^zS_{b,i}^z.
\end{array}
\end{equation}

The antiferromagnetic coupling $J>0$ describes the strength of the spin-spin interactions between
interstitial ionic dimers $\mathrm{Mn}^{2+}-\mathrm{Cu}^{2+}$, while dimer-monomer interaction $J_{0}$ stands for the Ising-type exchange coupling between nodal and interstitial magnetic ions $\mathrm{Fe}^{3+}-\mathrm{Mn}^{2+}$. $\Delta$ is the anisotropy parameter, and $B_{k}=g_k\mu_0B$ with $k=\{1,2,3\}$ is an external magnetic field along the $z$-axis (for simplicity we put $\mu_0=1$). $\mu_i$ and $\mu_{i+1}$, taking values $(\frac{1}{2},\;-\frac{1}{2})$, label the Ising nodal spins $\mathrm{Fe}^{3+}$.

The impurity induced  Hamiltonian is described as follow

\[
\begin{array}{cl}
\mathcal{H}_{i}^{imp}= J\left(\mathbf{S}_{a,i},\mathbf{S}_{b,i}\right)_{\Delta}+J_{0}S_{a,i}^{z}\left(\mu_{i}+\mu_{i+1}\right)
 -h_{2}S_{a,i}^{z}-h_{3}S_{b,i}^{z}-\frac{B_{1}}{2}\left(\mu_{i}+\mu_{i+1}\right),\\
 \\
 \mathrm{for}\:i=r.
\end{array}
\]
by assuming magnetic impurity $h_{k}=g_{k}B\left(1+\gamma\right)$ with $k=\{2,3\}$, for which $\gamma$ denotes the induced impurity parameter.

Four relevant eigenenergies per unit cell of the Hamiltonian $\mathcal{H}_{i}^{host}$ are quickly achieved by diagonalizing $i-$th XXZ
dimer. Thus, we have following expressions

\begin{equation} \label{eq:Host_energies}
\begin{array}{cl} 
\varepsilon_{1,4}= & \frac{J\Delta}{4}\pm\frac{J_{0}}{2}\left(\mu_{i}+\mu_{i+1}\right)-\frac{B_{1}}{2}\left(\mu_{i}+\mu_{i+1}\right)\mp\frac{1}{2}\left(B_{2}+B_{3}\right),\\
\varepsilon_{2,3}= & -\frac{J\Delta}{4}-\frac{B_{1}}{2}\left(\mu_{i}+\mu_{i+1}\right)\pm\sqrt{\Omega^{2}+J^{2}},
\end{array}
\end{equation}
where, $\Omega=J_{0}\left(\mu_{i}+\mu_{i+1}\right)-\left(B_{2}-B_{3}\right)$.

The corresponding eigenvectors to Eq. (\ref{eq:Host_energies}) in the standard dimer basis $\left\{ \left|00\right>,\left|01\right>,\left|10\right>,\left|11\right>\right\} $
are

\begin{eqnarray}
\left|\varphi_{1}\right> & = & \left|00\right>,\nonumber \\
\left|\varphi_{2}\right> & = & m_{+}\left|01\right>+n_{+}\left|10\right>,\nonumber \\
\left|\varphi_{3}\right> & = & m_{-}\left|01\right>+n_{-}\left|10\right>,\nonumber \\
\left|\varphi_{4}\right> & = & \left|11\right>.
\end{eqnarray}
where following notations are adopted
\begin{equation} 
\begin{array}{cl} 
 m_{\pm}=\frac{J}{\sqrt{2J^{2}+2\Omega^{2}\mp2\Omega\sqrt{\Omega^{2}+J^{2}}}},\quad
n_{\pm}=\frac{\Omega\pm\sqrt{\Omega^{2}+J^{2}}}{\sqrt{2J^{2}+2\Omega^{2}\pm2\Omega\sqrt{\Omega^{2}+J^{2}}}}.
\end{array}
\end{equation}
Here, $\mid 0\rangle$ and $\mid1\rangle$  reveal two eigenbasis of the spin operator $S_i^z$ with the respective eigenvalues
 $S_i^z =[\frac{1}{2},\;-\frac{1}{2}]$.
 
Analogously, eigenvalues of $\mathcal{H}_{i}^{imp}$ can be obtained from below formulae

\begin{equation} \label{eq:imp_energies}
\begin{array}{cl} 
\widetilde{\varepsilon}_{1,4}= \frac{J\Delta}{4}\pm\frac{J_{0}}{2}\left(\mu_{i}+\mu_{i+1}\right)-\frac{B_{1}}{2}\left(\mu_{i}+\mu_{i+1}\right)\mp\frac{1}{2}\left(h_{2}+h_{3}\right),\\
\widetilde{\varepsilon}_{2,3}= -\frac{J\Delta}{4}-\frac{B_{1}}{2}\left(\mu_{i}+\mu_{i+1}\right)\pm\sqrt{\kappa^{2}+J^{2}},
\end{array}
\end{equation}
where $\kappa=J_{0}\left(\mu_{i}+\mu_{i+1}\right)-\left(h_{2}-h_{3}\right)$.
The corresponding eigenstates are represented as

\begin{eqnarray}
\left|\widetilde{\varphi}_{1}\right> & = & \left|00\right>,\nonumber \\
\left|\widetilde{\varphi}_{2}\right> & = & \Sigma_{+}\left|01\right>+\Gamma_{+}\left|10\right>,\nonumber \\
\left|\widetilde{\varphi}_{3}\right> & = & \Sigma_{-}\left|01\right>+\Gamma_{-}\left|10\right>,\nonumber \\
\left|\widetilde{\varphi}_{4}\right> & = & \left|11\right>.
\end{eqnarray}
where
\begin{equation} \label{eq:GammaSigma}
\begin{array}{cl} 
 \Sigma_{\pm}=\frac{J}{\sqrt{2J^{2}+2\kappa^{2}\mp2\kappa\sqrt{\kappa^{2}+J^{2}}}},\quad
\Gamma_{\pm}=\frac{-\kappa\pm\sqrt{\kappa^{2}+J^{2}}}{\sqrt{2J^{2}+2\kappa^{2}\mp2\kappa\sqrt{\kappa^{2}+J^{2}}}}.
\end{array}
\end{equation}
The system state at thermal equilibrium is $\rho(T)=\frac{\exp(-\beta \mathcal{H})}{Z}$,
where $\beta=1/k_{B}T$, with $k_{B}$ being the Boltzmann's constant,
$T$ is the absolute temperature, whereas $Z=Tr\left[\exp(-\beta \mathcal{H})\right]$ is the partition function of the system.

\subsection{The density operator and partition function}

To gain an overall insight into the quantum properties of the magnetic $\mathrm{Mn}^{2+}-\mathrm{Cu}^{2+}$ ion dimers such as thermal entanglement, the quantum coherence, QFI and teleportation etc., it is necessary to obtain  partition function of the model under consideration.
Beforehand, the original version of this model has been exactly solved in the thermodynamic limit through usual transfer matrix technique \cite{souza}. The transfer matrix approach  would also be a good candidate to exactly solve such a model when an impurity dimer embedded in the  chain. Owing to this fact, the local density operator for Heisenberg dimer seating at site $i=r$ (site $a$ and $b$ labeled in Fig. \ref{fig:model})  whose magnetic $Mn^{2+}$ ion is bonded by two nodal Ising particles $\mu_i$ and $\mu_{i+1}$ reads

\begin{equation} \label{eq:Q_Operator}
\begin{array}{cl} 
\varrho(\mu_i,\;\mu_{i+1})=\sum\limits_{j=1}^4 e^{-\beta\varepsilon_{i,j}(\mu_i,\;\mu_{i+1})}\mid \varphi_{i,j}\rangle\langle\varphi_{i,j}\mid
\end{array}
\end{equation}
in which, $\varepsilon_{i,j}(\mu_i,\;\mu_{i+1})$ are two-qubit operator eigenvalues (\ref{eq:Host_energies}).
In the one hand, the partition function of the favorite model can be characterized in terms of Boltzmann factor of the $i-$ block dimer as following
\begin{equation} \label{eq:Bolt_Facts}
\begin{array}{cl} 
w(\mu_i,\;\mu_{i+1})=\mathrm{tr}_{ab}\big[\varrho(\mu_i,\;\mu_{i+1})\big]=\sum\limits_{j=1}^4 e^{-\beta\varepsilon_{i,j}(\mu_i,\;\mu_{i+1})}.
\end{array}
\end{equation}
As our system includes a dimer impurity at site $i=r$, gaining the Boltzmann factor of the impurity is momentous to generate total partition function. So, on the other hand, the Boltzmann factor for an embedded impurity could be written  by

\[
\widetilde{w}(\mu_{i},\mu_{i+1})=\sum_{j=1}^{4}\mathrm{e}^{-\beta\widetilde{\mathcal{E}}_{ij}(\mu_{i},\mu_{i+1})}\;.
\]
The model partition function can be thus written in terms of the multiplication of Boltzmann factors, i.e.,
\begin{equation}
\begin{array}{cl} Z_{N}= & \sum_{\{\mu\}}w(\mu_{1},\mu_{2})\ldots w(\mu_{r-1},\mu_{r})\widetilde{w}(\mu_{r},\mu_{r+1})
  w(\mu_{r+1},\mu_{r+2})\ldots w(\mu_{N},\mu_{1})\;.\label{eq:rho-df-1}
\end{array}
\end{equation}
Above equation can be straightforwardly defined  by 
\[
 Z_{N}=\mathrm{tr}\left(\widetilde{W}W^{N-1}\right),
 \]
 in which the transfer-matrix $\widetilde{W}$ associated to the whole of spin chain except the impurity part reads
 
\begin{equation}
W=\left[\begin{array}{cc}
w(\frac{1}{2},\frac{1}{2}) & w(\frac{1}{2},-\frac{1}{2})\\
w(-\frac{1}{2},\frac{1}{2}) & w(-\frac{1}{2},-\frac{1}{2})
\end{array}\right]\;.\label{eq:W}
\end{equation}
In a similar way, the transfer-matrix $\widetilde{W}$ for the impurity is given by
\[
\widetilde{W}=\left[\begin{array}{cc}
\widetilde{w}(\frac{1}{2},\frac{1}{2}) & \widetilde{w}(\frac{1}{2},-\frac{1}{2})\\
\widetilde{w}(-\frac{1}{2},\frac{1}{2}) & \widetilde{w}(-\frac{1}{2},-\frac{1}{2})
\end{array}\right]\;.
\]
The transfer-matrix elements adopts following notations
 $w_{++}\equiv w(\frac{1}{2},\frac{1}{2})$, $w_{+-}\equiv w(\frac{1}{2},-\frac{1}{2})$, $w_{-+}\equiv w(-\frac{1}{2},\frac{1}{2})$ and $w_{--}\equiv w(-\frac{1}{2},-\frac{1}{2})$.
Accordingly, the eigenvalues of the aforedescribed transfer-matrix $W$ are given by,
\begin{equation}
\Lambda_{\pm}=\frac{w_{++}+w_{--}\pm Q}{2}\;,
\end{equation}
assuming $Q=\sqrt{\left(w_{++}-w_{--}\right)^{2}+4w_{+-}^{2}}$.
Therefore,  under periodic boundary conditions, the total partition function for a finite-size spin chain analogous to our model is given by
\begin{equation}
Z_{N}=a\Lambda_{+}^{N-1}+d\Lambda_{-}^{N-1}\;,
\end{equation}
where 
\[
\begin{array}{cc}
a= & \frac{4w_{+-}\widetilde{w}_{+-}+\left(w_{++}-w_{--}\right)\left(\widetilde{w}_{++}-\widetilde{w}_{--}\right)+Q\left(\widetilde{w}_{++}+\widetilde{w}_{--}\right)}{2Q},\\
\\
d= & \frac{-4w_{+-}\widetilde{w}_{+-}-\left(w_{++}-w_{--}\right)\left(\widetilde{w}_{++}-\widetilde{w}_{--}\right)+Q\left(\widetilde{w}_{++}+\widetilde{w}_{--}\right)}{2Q}\;.
\end{array}
\]
In the thermodynamic limit $N\rightarrow\infty$, the partition function is routinely determined by the largest eigenvalue of the transfer matrix $W$ as we labeled above $\Lambda_+$. Hence, we attain $Z_{N}=a\Lambda_{+}^{N-1}$. Now, we are able to investigate the thermal quantum correlations, the concurrence, QFI and quantum teleportation after obtaining the reduced density operator $\widetilde{\rho}$ of the dimer impurity.
%magnetic $\mathrm{Mn}^{2+}-\mathrm{Cu}^{2+}$ ions.
%

\subsection{ Two-qubit density operator in a matrix form}
At this stage, we employ the approach reported in Refs. \cite{mo-1,mo-2} in order to calculate the thermal average of the two-qubit operator corresponding to an impurity linked by Ising nodal ions $\mu_r$ and $\mu_{r+1}$. This operator in the standard  dimer basis becomes
\begin{equation} \label{eq:varrho}
\begin{array}{cl} 
\widetilde{\varrho}(\mu_{r},\mu_{r+1})=\left[\begin{array}{cccc}
\widetilde{\varrho}_{1,1} & 0 & 0 & 0\\
0 & \widetilde{\varrho}_{2,2} & \widetilde{\varrho}_{2,3} & 0\\
0 & \widetilde{\varrho}_{3,2} & \widetilde{\varrho}_{3,3} & 0\\
0 & 0 & 0 & \widetilde{\varrho}_{4,4}
\end{array}\right],
\end{array}
\end{equation}
with below expressions

\begin{equation*}
\begin{array}{cl} 
\widetilde{\varrho}_{1,1}(\mu_{r},\mu_{r+1})= & \mathrm{e}^{-\beta\widetilde{\varepsilon}_{r1}},\\
\widetilde{\varrho}_{2,2}(\mu_{r},\mu_{r+1})= & \mathrm{e}^{-\beta\widetilde{\varepsilon}_{r2}}\Sigma_{+}^{2}+\mathrm{e}^{-\beta\widetilde{\varepsilon}_{r3}}\Sigma_{-}^{2},\\
\widetilde{\varrho}_{2,3}(\mu_{r},\mu_{r+1})= & \widetilde{\varrho}_{3,2}(\mu_{r},\mu_{r+1})= \mathrm{e}^{-\beta\widetilde{\varepsilon}_{r2}}\Sigma_{+}\Gamma_{+}+\mathrm{e}^{-\beta\widetilde{\varepsilon}_{r3}}\Sigma_{-}\Gamma_{-},\\
\widetilde{\varrho}_{3,3}(\mu_{r},\mu_{r+1})= & \mathrm{e}^{-\beta\widetilde{\varepsilon}_{r2}}\Gamma_{+}^{2}+\mathrm{e}^{-\beta\widetilde{\varepsilon}_{r3}}\Gamma_{-}^{2},\\
\widetilde{\varrho}_{4,4}(\mu_{r},\mu_{r+1})= & \mathrm{e}^{-\beta\widetilde{\varepsilon}_{r4}}.
\end{array}
\end{equation*}
Four coefficients $\Gamma_{\pm}$ and $\Sigma_{\pm}$ have already defined in Eq. (\ref{eq:GammaSigma}). 

\subsection{The reduced density state for the embedded impurity}
The reduced density matrix of the dimer impurity at thermal equilibrium can be eventually expressed as the X state

\begin{equation}\label{eq:rhoT}
\widetilde{\rho}(T)=\left[\begin{array}{cccc}
\widetilde{\rho}_{11} & 0 & 0 & 0\\
0 & \widetilde{\rho}_{22} & \widetilde{\rho}_{23} & 0\\
0 & \widetilde{\rho}_{23} & \widetilde{\rho}_{33} & 0\\
0 & 0 & 0 & \widetilde{\rho}_{44}
\end{array}\right],
\end{equation}
whose elements after performing below time-consuming analysis are achieved.

The components $\widetilde{\rho}_{k,l}$ can be identified by
\begin{equation}
\widetilde{\rho}_{k,l}= \frac{1}{Z_{N}}\sum_{\{\mu\}}w(\mu_{1},\mu_{2})\ldots w(\mu_{r-1},\mu_{r})\widetilde{\varrho}_{k,l}(\mu_{r},\mu_{r+1})
 w(\mu_{r+1},\mu_{r+2})\ldots w(\mu_{N},\mu_{1})\;.\label{eq:rho-df}
\end{equation}
The transfer-matrix technique results in reforming elements $\widetilde{\rho}_{k,l}$ as
\begin{equation}
\tilde{\rho}_{k,l}=\frac{1}{Z_{N}}\mathrm{tr}\left(W^{r-1}\widetilde{P}_{k,l}W^{N-r}\right)=\frac{1}{Z_{N}}\mathrm{tr}\left(\widetilde{P}_{k,l}W^{N-1}\right)\;,
\end{equation}
where
\begin{equation}
\widetilde{P}_{k,l}=\left[\begin{array}{cc}
\widetilde{\varrho}_{k,l}(\frac{1}{2},\frac{1}{2}) & \widetilde{\varrho}_{k,l}(\frac{1}{2},-\frac{1}{2})\\
\widetilde{\varrho}_{k,l}(-\frac{1}{2},\frac{1}{2}) & \widetilde{\varrho}_{k,l}(-\frac{1}{2},-\frac{1}{2})
\end{array}\right],
\end{equation}
by hypothesizing  $\widetilde{\varrho}_{k,l}(++)\equiv\widetilde{\varrho}_{k,l}(\frac{1}{2},\frac{1}{2})$,
$\widetilde{\varrho}_{k,l}(+-)\equiv\widetilde{\varrho}_{k,l}(\frac{1}{2},-\frac{1}{2})$,
$\widetilde{\varrho}_{k,l}(-+)\equiv\widetilde{\varrho}_{k,l}(-\frac{1}{2},\frac{1}{2})$,
$\widetilde{\varrho}_{k,l}(--)\equiv\widetilde{\varrho}_{k,l}(-\frac{1}{2},-\frac{1}{2})$. 
To accomplish our analytical procedure to deduce the special components of the partially averaged reduced density matrix 
$\widetilde{\rho}_{k,l}$ we proceed to reproduce the unitary transformation operator $U$ that diagonalizes the transfer matrix $W$.
This operator is given by 
\begin{equation}
U=\left[\begin{array}{cc}
\Lambda_{+}-w_{--} & \Lambda_{-}-w_{--}\\
w_{+-} & w_{+-}
\end{array}\right],
\end{equation}
such that its inverse matrix can be defined by
\begin{equation}
U^{-1}=\left[\begin{array}{cc}
\frac{1}{Q} & -\frac{\Lambda_{-}-w_{--}}{Qw_{+-}}\\
-\frac{1}{Q} & \frac{\Lambda_{+}-w_{--}}{Qw_{+-}}
\end{array}\right]\;.
\end{equation}
Consequently, the special components of the two-qubit density operator of the dimer impurity at site $i=r$ is specified by
\begin{equation}
\widetilde{\rho}_{k,l}=\frac{\mathrm{tr}\left(U^{-1}\widetilde{P}_{k,l}U\left[\begin{array}{cc}\Lambda_{+}^{N-1} & 0\\
0 & \Lambda_{-}^{N-1}
\end{array}\right]\right)}{Z_N}\;.
\end{equation}
If one calculates all elements $\widetilde{\rho}_{k,l}$ out of curiosity, in the thermodynamic limit, will find following relation
\begin{equation*}
\widetilde{\rho}_{k,l}= \frac{\mathcal{A}_{k,l}+\mathcal{B}_{k,l}}{\mathcal{M}}\;,
\end{equation*}
where 
\[
\begin{array}{cl}
\mathcal{A}_{k,l}= & Q\left[\widetilde{\varrho}_{k,l}(++)+\widetilde{\varrho}_{k,l}(--)\right]+4w_{+-}\widetilde{\varrho}_{k,l}(+-),\\
\mathcal{B}_{k,l}= & \left[\widetilde{\varrho}_{k,l}(++)-\widetilde{\varrho}_{k,l}(--)\right]\left(w_{++}-w_{--}\right),\\
\mathcal{M}= & Q\left(\widetilde{w}_{++}+\widetilde{w}_{--}\right)+4w_{+-}\widetilde{w}_{+-}+\\
 & +\left(\widetilde{w}_{++}-\widetilde{w}_{--}\right)\left(w_{++}-w_{--}\right)\;.
\end{array}
\]

Now, quantum correlations, thermal entanglement,  the $l_{1}$-norm of coherence $\mathcal{C}_{l_{1}}$, QFI and  the average fidelity $F_{A}$ for the impurity part of the model can be simply investigated using above achievements.

\section{Quantum Correlations}\label{QC}

In this section, we discuss correlations between the magnetic ions of the dimer impurity $\mathrm{Mn}^{2+}-\mathrm{Cu}^{2+}$ induced in the chain as local spin-spin quantum correlations.  Mainly for the sake of commutation relation $[H,\;\sum_{i=1}^NS^z_i]=0$ and the translation invariance, the two point spin-spin correlation functions corresponding to the $x$ and the $z$ axis can be characterized in terms of the reduced density operator elements 
\cite{Werlang2010}, namely,
\begin{equation} \label{eq:varrho}
\begin{array}{cl} 
 \langle S^x_{Mn}S^x_{Cu}\rangle=\frac{\widetilde{\rho}_{22}}{2},\quad
 \langle S^z_{Mn}S^z_{Cu}\rangle =\frac{1}{4}-\widetilde{\rho}_{23}
\end{array}
\end{equation}

\subsection{Thermal entanglement }
The two-qubit reduced density matrix $\widetilde{\rho}_{k,l}$ provides all thing needed to analyze the bipartite entanglement.
A good measure of the thermal entanglement between pair spins of impurity  is provided by the concurrence and shall be quantified by the 
 $\mathcal{C}$ as \cite{wootters}
\begin{eqnarray}  
\mathcal{C}={\rm {max}\left\{ 0,{\lambda_{1}}-{\lambda_{2}}-{\lambda_{3}}-{\lambda_{4}}\right\}},\label{eq:conco}
\end{eqnarray}
where $\lambda_{i}\:(i=1,2,3,4)$ are the square root of the eigenvalues of the matrix $R=\rho\left(\sigma^{y}\otimes\sigma^{y}\right)\rho^{\ast}\left(\sigma^{y}\otimes\sigma^{y}\right)$ in descending order (with $\sigma^{y}$ being the Pauli matrix). 

For  the system in thermal equilibrium with density matrix in the X-form (\ref{eq:rhoT}) the concurrence is squarely given by
\[
\mathcal{C}(\widetilde{\rho})=2\mathrm{max}\{|\widetilde{\rho}_{2,3}|-\sqrt{\widetilde{\rho}_{1,1}\widetilde{\rho}_{4,4}},0\}\;.
\]

Next, we aim to examine the influences of the magnetic impurity embedded in the  Ising-XXZ chain under consideration on the thermal entanglement of the Heisenberg dimer at site $i=r$, on the quantum coherence, on the QFI, further on the quantum teleportation. 

In plots correspond to the concurrence, $l_{1}$-norm of coherence and average fidelity, solid lines show information about the original model (without impurity), whereas dashed lines indicate information for the model consisting of a static dimer impurity with $\gamma=-0.8$. 
In all forthcoming figures, we consider the specific values of the gyromagnetic factors $g_{1}=1.2$, $g_{2}=5$, $g_{3}=1.1$ associated to the magnetic ions $\mathrm{Fe}^{3+}$, $\mathrm{Mn}^{2+}$ and $\mathrm{Cu}^{2+}$, respectively. Besides, in our numerical computations the Heisenberg coupling will be selected as the energy unit, i.e., $|J| = 1$.

\begin{figure}
\includegraphics[scale=0.6]{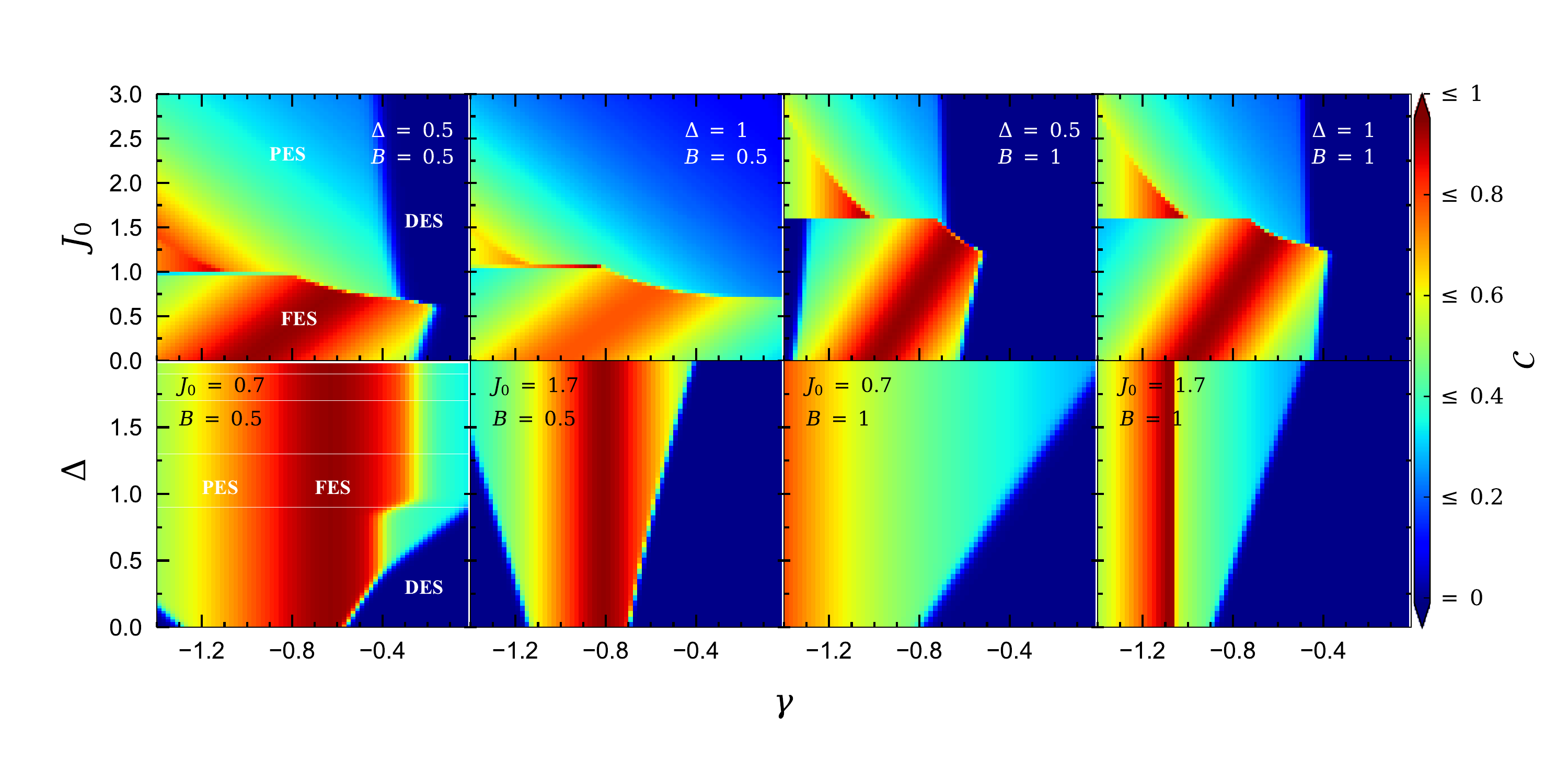}
\caption{ \label{Fig2}  Upper row panels illustrate the concurrence $\mathcal{C}$ in the $(\gamma-J_0)$  plane at low temperature $T=0.02$, where for each of exchange anisotropy and magnetic field  parameters two values $\Delta=\{0.5,\;1\}$  and $B=\{0.5,\;1\}$ are considered. Panels in lower row depict the concurrence in the  $(\gamma-\Delta)$  plane such that  $J_0=\{0.7,\;1.7\}$ and $B=\{0.5,\;1\}$ are assumed.}
\end{figure}

In Fig. \ref{Fig2}  (upper row panels) we display  the concurrence $\mathcal{C}$ in the 
$(J_0-\gamma)$ plane for two different values of  $B$ and $\Delta$ at finite low-temperature $T=0.02$.
In this figure, three different phases: disentangled state (DES), fully entangled state (FES) with an oblique band region, and partially entangled state (PES) are observable (see first panel). By tuning the both parameters $B$ and $\Delta$,  phase boundaries undergo substantial changes.
Similar to this, in lower row panels, we depict finite low-temperature $\Delta$-dependence of the concurrence versus the impurity parameter
 $\gamma$ for two different values of $B$ and $J_0$. It could be understood that at a specific rage of $\gamma$ there is a fully entangled state band which is independent of the anisotropy $\Delta$. Increasing exchange coupling $J_0$ leads to strength the DES regime and to limit FES boundary. 

\begin{figure}
\includegraphics[scale=0.6]{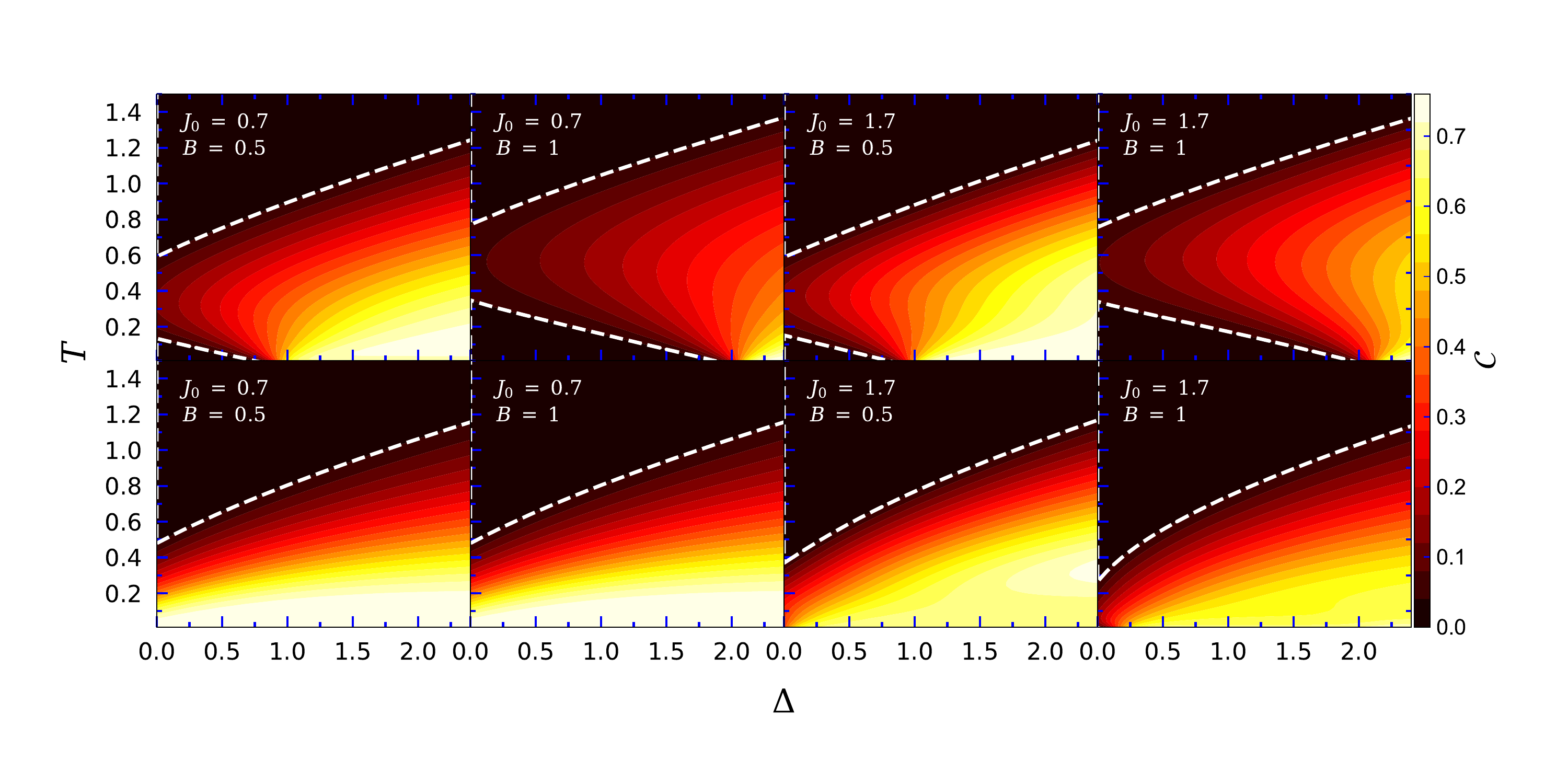}
\caption{ \label{Fig22}  The density plot of the concurrence in the $(\Delta-T)$ plane, where exchange coupling $J_0$ and magnetic field $B$  have been substitutionally taken up as $J_0=\{0.7,\;1.7\}$  and $B=\{0.5,\;1\}$. Panels plotted in upper row represent the concurrence for the original model without impurity ($\gamma=0$), whereas plots in lower row show the concurrence for the model with magnetic impurity
  ($\gamma=-0.8$) at site $i=r$. Dashed lines demonstrate the threshold temperature (at which the entanglement death occurs) against the anisotropy $\Delta$.}
\end{figure}

To understand the anisotropy dependence of the threshold temperature at which the entanglement vanishes, we have plotted in Fig. \ref{Fig22}
the density plot of the concurrence in the $\Delta-T$ plane such that the threshold temperature behavior versus $\Delta$ is engraved by dashed line in each panel. Plots in upper row present the concurrence for the original model  ($\gamma=0$), while lower row panels represent the same term but for the model with a magnetic impurity, supposing  $\gamma=-0.8$.  For the original mode, one sees there are two threshold temperature points which shifts toward higher temperatures upon increasing both of $B$ and $J_0$. Besides, the entanglement death happens for higher anisotropies (see second and fourth panels). But on the other hand, for the case when the model involves with a magnetic impurity at site $i=r$ there is a single threshold temperature for each value of exchange anisotropy $\Delta$, revealing the model with a magnetic impurity does not show re-entrant threshold temperature phenomenon.

\begin{figure}
\centering
\includegraphics[scale=0.3]{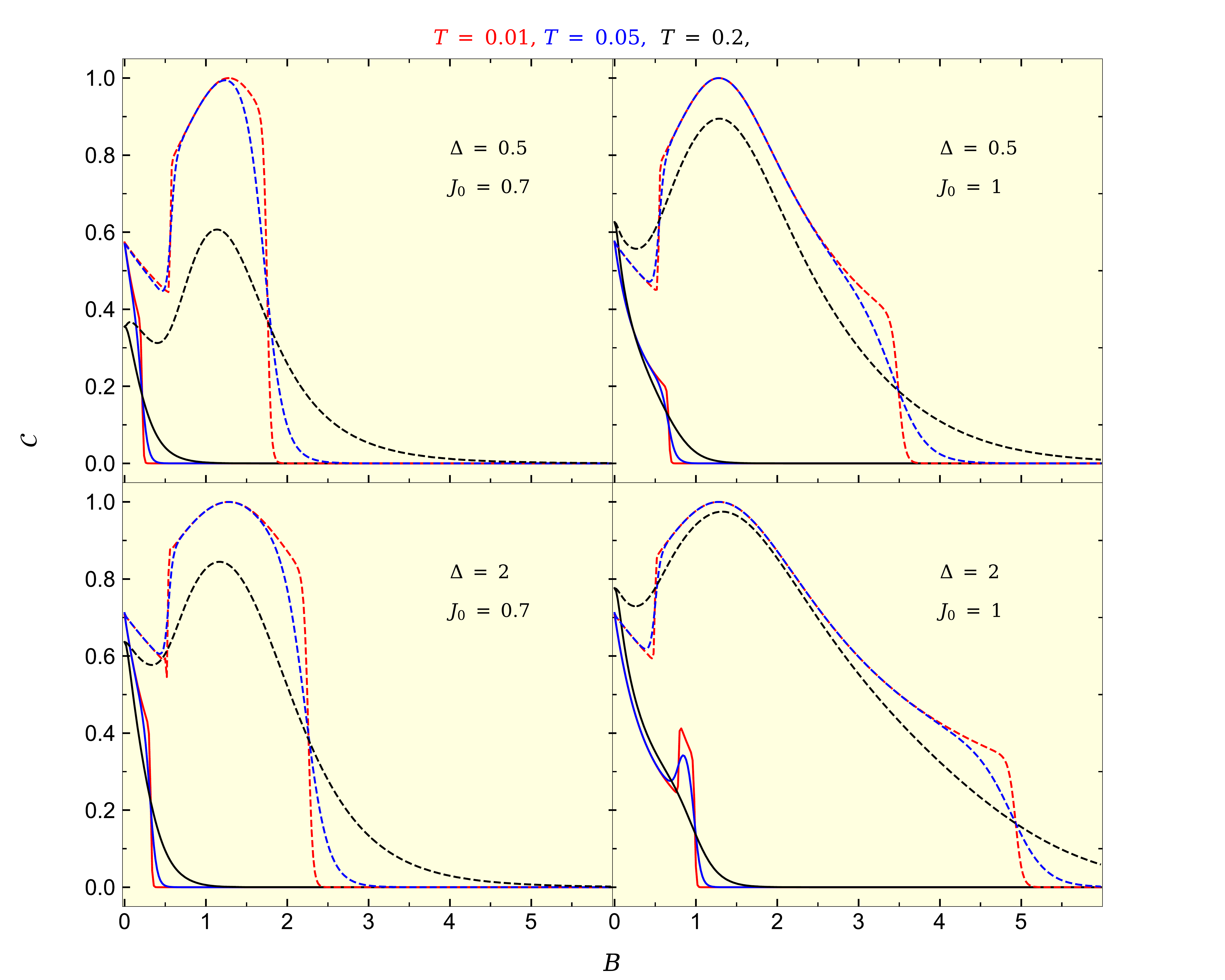}
\caption{\label{Fig3} The concurrence $\mathcal{C}$ as a function of the magnetic field $B$, assuming three different temperatures   $T=0.01$ (red lines), $T=0.05$ (blue lines) and $T=0.2$ (black lines) and impurity parameter $\gamma=-0.8$. g-factors have been taken as $g_{1}=1.2$,
$g_{2}=5$, $g_{3}=1.1$. Totally, we consider two different values of $\Delta=\{0.5,\;2\}$, increasing from up to down panels, and $J_0=\{0.7,\;1\}$, increasing from  left to right panels.}
\end{figure}

Regarding above findings, in Fig. \ref{Fig3} is shown the concurrence $\mathcal{C}$ as a function of magnetic field $B$ for three different  temperatures $T=0.01$ (red lines), $T=0.05$ (blue lines) and $T=0.2$ (black lines), where $J_{0}$ and $\Delta$ have been considered to be tuneable terms. From left to right  the figure shows the concurrence versus magnetic field when all parameters set to be fixed values except $J_0$. From up to down is considered all parameters as fixed values except $\Delta$.
% Solid lines of this figure corresponding to the original model and those plotted in Ref. \cite{Str2019} are the same. 
As reported in previous work, at finite low temperature and low magnetic field the state of $\mathrm{Mn}^{2+}-\mathrm{Cu}^{2+}$ without magnetic impurity is partially entangled. With increase of the temperature the concurrence remarkably decreases. 
By embedding a magnetic impurity to the site $r$, we see a significant change in concurrence behavior. Surprisingly, the concurrence of the dimer impurity is alive for higher magnetic field while for the original model it vanishes. When the anisotropy $\Delta$ increases, the concurrence increases and reaches its maximum value $\mathcal{C}=1$ at a special point on the magnetic field axis. With increase of the exchange coupling $J_0$,
the concurrence increases for the both considered modes with and without magnetic impurity. But, the influence of $J_0$ on the impurity case is much more sensible than without impurity mode. Meanwhile, the sharp maximum of the concurrence curve changes to a dome-shaped maximum, revealing the fact that the entanglement death occurs for the remarkably higher magnetic fields.

 %One sees a threshold temperature at which the concurrence vanishes. 
%Dashed lines depict the concurrence of the dimer impurity for the same parameter sets to solid lines (we consider the impurity parameter to be fixed $\gamma=-0.8$). 

The most interesting finding from this special consideration is that, when the model is putted in an external uniform magnetic field (the system without the impurity), the state of the Heisenberg dimers does not reach maximum entanglement ($\mathcal{C}=1$) even for low magnetic fields. When the model possesses an impurity the situation completely changes (dashed lines plotted in Fig.  \ref{Fig3}), namely, by increasing the magnetic field the concurrence in the magnetic dimer with an impurity $\gamma=-0.8$ goes to become maximally entangled. Surprisingly, the state of the model becomes maximally entangled at low temperature and critical  magnetic field $B_{max}=1.282$. With further increase of the magnetic field the concurrence decreases and the state of the model with impurity losses its maximum entanglement property. Another notable remark from this figure is that, for the case when the model does not encompass the impurity, the threshold magnetic field at which the concurrence vanishes, moves toward higher values as the both of parameters $\Delta$ and $J_0$ increase monotonically.

\begin{figure}
\includegraphics[scale=0.3]{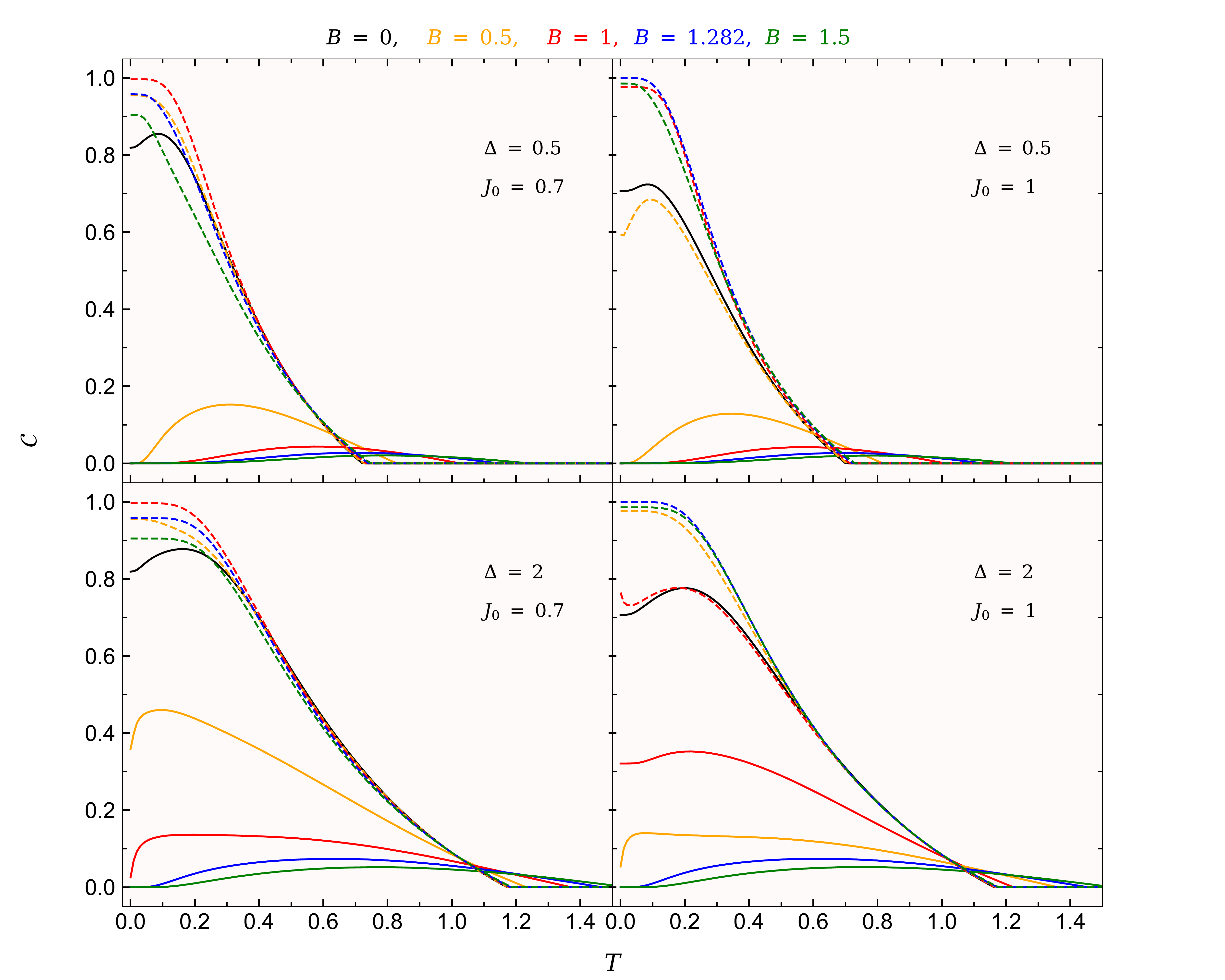}
\caption{\label{Fig4} The concurrence $\mathcal{C}$ as a function of temperature 
$T$, for several fixed values of the magnetic field, assuming the same set of other parameters to Fig. \ref{Fig3}.}
\end{figure} 

In Fig. \ref{Fig4} is illustrated the temperature dependence of concurrence $\mathcal{C}$  for several values of the magnetic field, where parameters $\Delta$ and $J_0$ are again assumed to be tuneable terms. In this case, we see that by inducing a magnetic impurity to the Ising-Heisenberg heterotrimetallic chain, the concurrence substantially increases at low temperatures. For the original model, the concurrence does not reach maximally entangled state even for the low magnetic fields. It is quite clear that for the dimer impurity at low magnetic fields, the concurrence reaches its maximum $\mathcal{C}=1$ as the temperature decreases. Increasing both parameters $\Delta$ and $J_0$ results in reaching maximum value of the concurrence at higher temperatures and higher magnetic fields (trace for example  evolution of the dashed red lines and dashed blue lines in all panels). Another unconventional phenomenon that is visible from this figure is that, for the original model the threshold temperature at which the concurrence vanishes is strongly dependent on the magnetic field, while for the embedded impurity model  the threshold temperature is almost a fixed constant when the magnetic field changes.
%  until applying associated magnetic field at which the system state becomes highly entangled.  

\subsection{Quantum Coherence}
To quantify the quantum coherence in a spin system a  trace-distance measure of coherence is adopted.
Naturally, we take up the $l_{1}$-norm of coherence that could be defined as
%The quantum coherence is useful resource for quantum information processing
%task. The quantum coherence in a bipartite system can be contained
%either locally or in the correlations between the subsystems. The
%portion of quantum coherence where all the coherence in the system
%is stored entirely within the quantum correlations which is called. 

\[
\mathcal{C}_{l_{1}}(\rho)=\sum_{i\neq j}|\langle i|\rho|j\rangle|.
\]

Above equation means that the $l_{1}$-norm of coherence is given by the sum of absolute values of all off-diagonal elements in the density matrix $\rho$.
\begin{figure}
\includegraphics[scale=0.45]{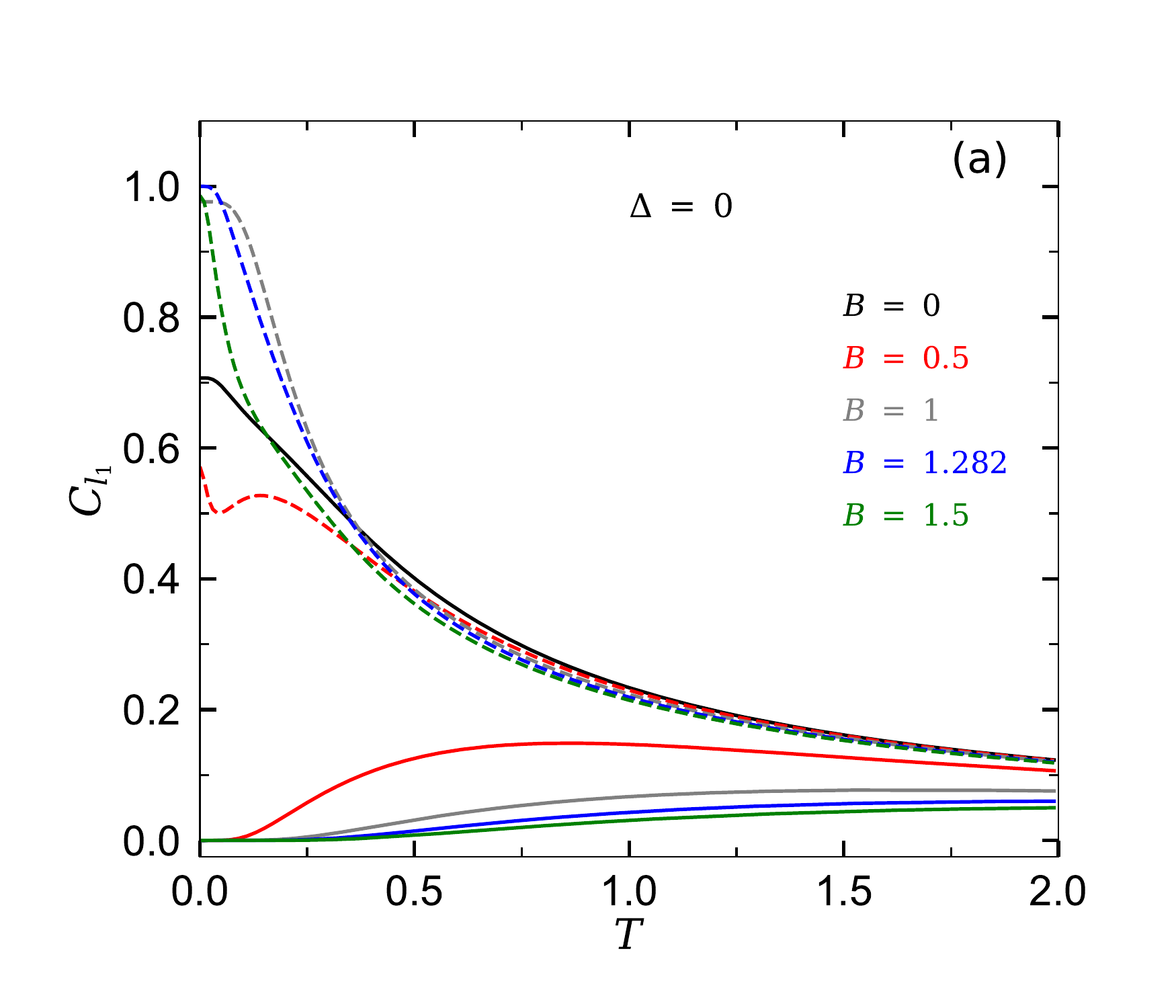}
\includegraphics[scale=0.45]{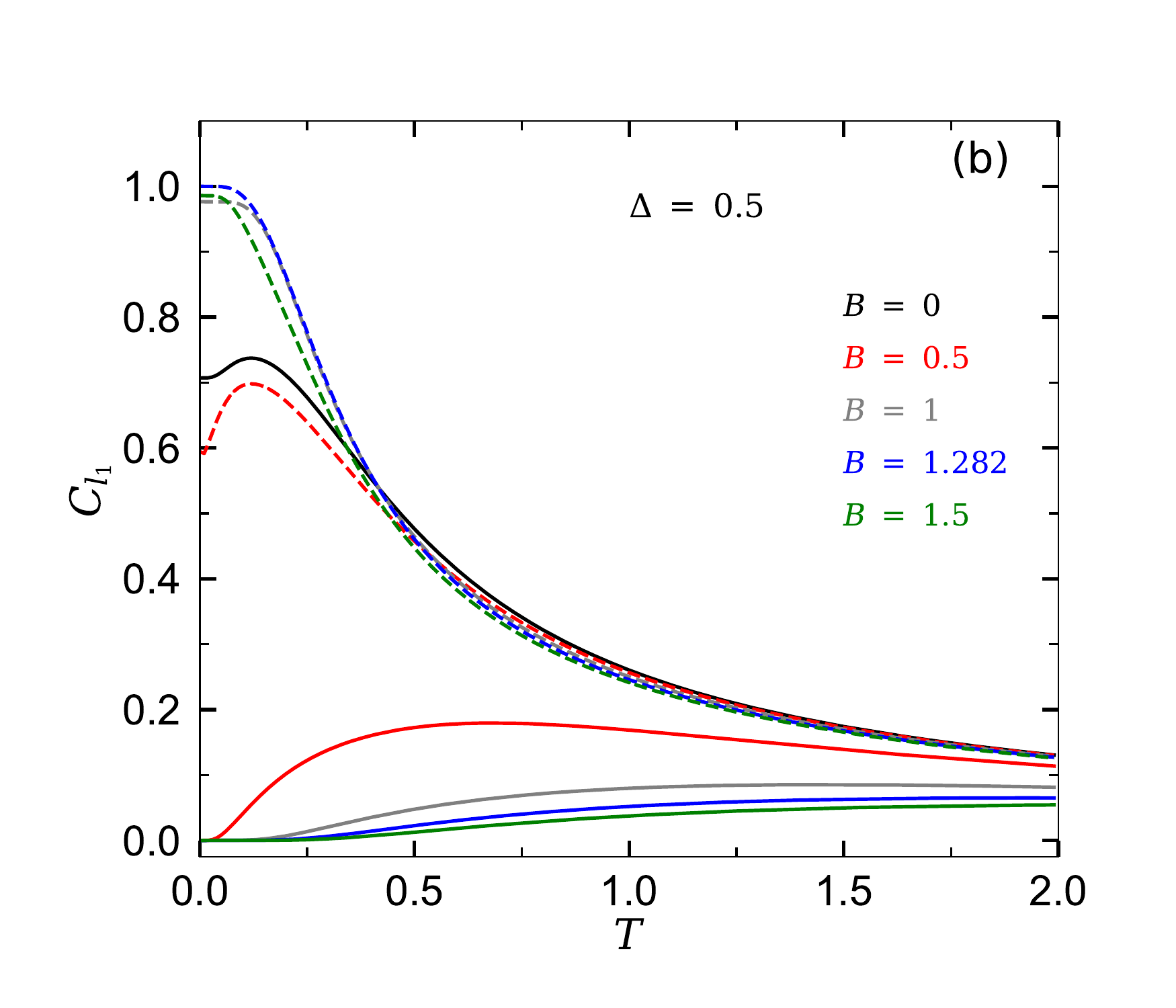}
\includegraphics[scale=0.45]{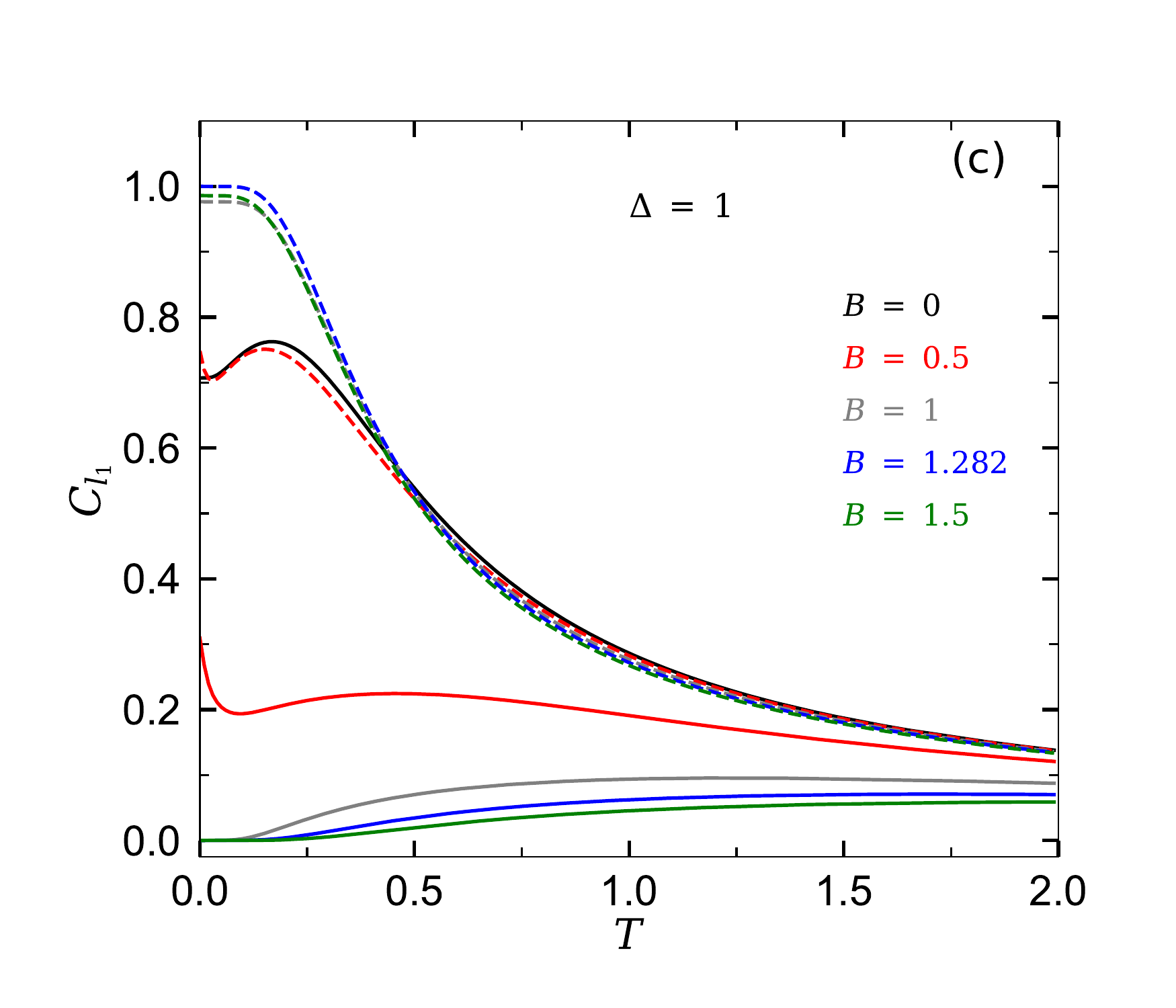}
\includegraphics[scale=0.45]{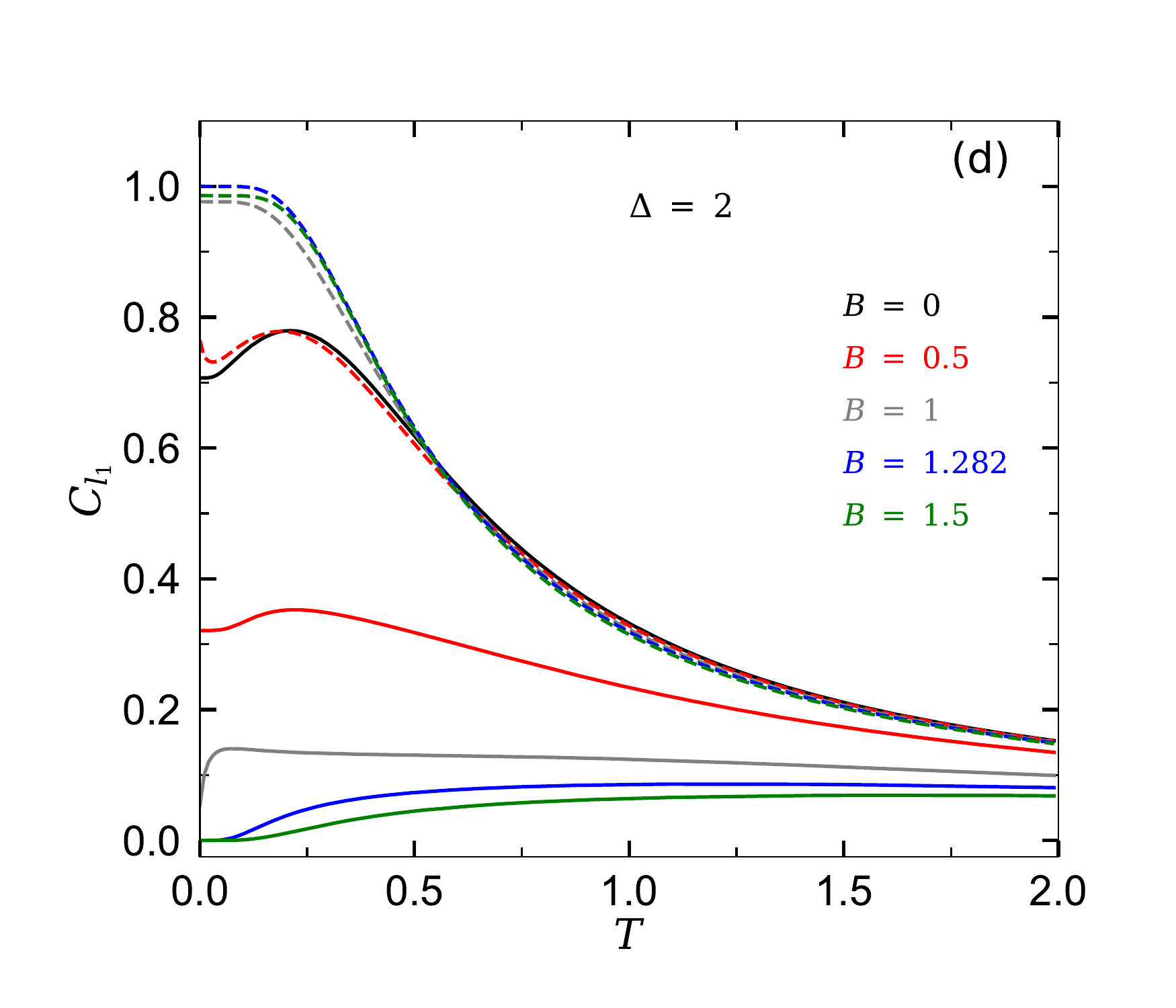}
\caption{\label{Fig5} The quantum coherence $\mathcal{C}_{l_{1}}$ as a function of temperature $T$ for several fixed values of the magnetic field and parameter set (a) $\Delta=0$, (b) $\Delta=0.5$, (c) $\Delta=1$, (d) $\Delta=2$, assuming fixed $J_0=1$ and the same set of other parameters to Fig. \ref{Fig3}.}
\end{figure}
We plot in Figs. \ref{Fig5}  (a)-(d) $l_{1}$-norm $\mathcal{C}_{l_{1}}$ as a function of the temperature for the same fixed magnetic fields to Fig. \ref{Fig4}, where  $J_0=1$ has been optionally hypothesized. An interesting result obtained for the quantum property of the model is belong to our finding for the temperature dependence of  $\mathcal{C}_{l_{1}}$.  In Fig. \ref{Fig5} (a) is displayed the  $l_{1}$-norm of coherence $\mathcal{C}_{l_{1}}$ versus temperature for the parameter set $\Delta=0$ and $J_0=1$, where several fixed values of the magnetic field have been selected. At low-temperature region, $l_{1}$-norm behavior against the temperature $T$ for the case when the model has a magnetic impurity is completely different from the original model. To clarify this point, under the cooling, norm $\mathcal{C}_{l_{1}}$ of the original model gradually decreases and eventually vanishes at a critical temperature (solid lines), demonstrating the system state becomes decoherence. While its curve intrinsically has a steep increase when an impurity induces to the model (see dashed lines). For the later, $l_{1}$-norm  reaches its maximum $\mathcal{C}_{l_{1}}=1$ at very low temperatures and threshold magnetic field $B=1.282$ (blue dashed line). With increase of the anisotropy $\Delta$, the low-temperature behavior of $\mathcal{C}_{l_{1}}$ undergoes substantial changes. For example, for the original  model, when anisotropic XXZ interaction is considered for the dimers ($\Delta>0$), the critical temperature at which $\mathcal{C}_{l_{1}}$ becomes zero shifts toward lower values (see panel Fig. \ref{Fig5}  (b)). When the anisotropy increases further, under cooling the norm $\mathcal{C}_{l_{1}}$ not only does not decreases but also increases monotonically (see Figs. \ref{Fig5}  (c) and (d)). 
For the model with impurity, as $\Delta$ increases, quantum coherence $\mathcal{C}_{l_{1}}$ reaches its maximum value at higher temperatures (see blue dashed line plotted in Fig. \ref{Fig5}  (d)).

\subsection{Quantum Fisher Information (QFI)}
Now, let us verify QFI and its first magnetic field derivative for the Ising-Heisenberg spin chain of the heterotrimetallic coordination compound 
$\mathrm{Fe-Mn-Cu}$ including magnetic impurity, and compare the QFI behavior with that of for the original model (without impurity).
In general phase estimation scenarios, the evolution of a quantum state, given by the reduced two-spin density matrix $\tilde{\rho}$ in Eq. (\ref{eq:rhoT}) with special components $\tilde{\rho}_{k,l}$, under a unitary transformation can be described as $\tilde{\rho}(\theta)=\exp[-i\mathcal{A}\theta]\tilde{\rho}\exp[i\mathcal{A}\theta]$, where $\theta$ is the phase shift and $\mathcal{A}$ is an operator. The estimation accuracy for $\theta$ is limited by the quantum
Cram{\' e}r-Rao inequality $\Delta\hat{\theta}\geq \frac{1}{\sqrt{\nu\mathcal{F}(\tilde{\rho}_{\theta})}}$.
where $\hat{\theta}$ expresses the unbiased estimator for $\theta$, and $\nu$ is the
number of times the measurement is repeated. Accordingly, the term $\mathcal{F}(\tilde{\rho}_{\theta})$ characterizes the QFI,
and is defined by \cite{Holevo,Liu2013}
\begin{equation} \label{eq:QFI}
\begin{array}{cl} 
 \mathcal{F}(\tilde{\rho}, \mathcal{A})=2\sum\limits_{i,j=1}^4\frac{(\tau_i-\tau_j)^2}{\tau_i+\tau_j}
 \mid\langle \chi_i\mid\mathcal{A}\mid\chi_j\rangle\mid^2,
\end{array}
\end{equation}
where $\vert\chi_i\rangle$ and $\tau_i$ are, respectively, the eigenvectors and their corresponding eigenvalues of the  density matrix $\tilde{\rho}$, and are utilized as the gauge states of estimation parameter $\theta$.
Since, we are going to examine the QFI for a typical Heisenberg dimer $\mathrm{Mn^{2+}-Cu^{2+}}$, along side of the density matrix $\tilde{\rho}$, we have to apply arbitrary complete sets of local orthonormal observables $\{\mathcal{A}_{\eta}\}$ and $\{\mathcal{B}_{\eta}\}$ associated to the both of subsystems $\mathrm{Mn}^{2+}$ and $\mathrm{Cu}^{2+}$, respectively. 
Correspondingly, the QFI for a general bipartite spin-1/2 system reads \cite{Li2013}
\begin{equation} \label{eq:QFI_2}
\begin{array}{cl} 
 \mathcal{F}=\sum\limits_{\eta} \mathcal{F}(\tilde{\rho}, \mathcal{A}_{\eta}\otimes I + I\otimes \mathcal{B}_{\eta}).
\end{array}
\end{equation}
In above, the local orthonormal observables $\{\mathcal{A}_{\eta}\}$ and $\{\mathcal{B}_{\eta}\}$ can be written as
\begin{equation} \label{eq:AB}
\begin{array}{cl} 
 \{\mathcal{A}_{\eta}\}=\{\mathcal{B}_{\eta}\}=\sqrt{2}\{I,\;S^x,\;S^y,\;S^z\},
\end{array}
\end{equation}
which $I$ is identity $2\times 2$ matrix.
Putting  this equation in Eq. (\ref{eq:QFI}), QFI $\mathcal{F}$ can be eventually evaluated.
\begin{figure}
\includegraphics[scale=0.45]{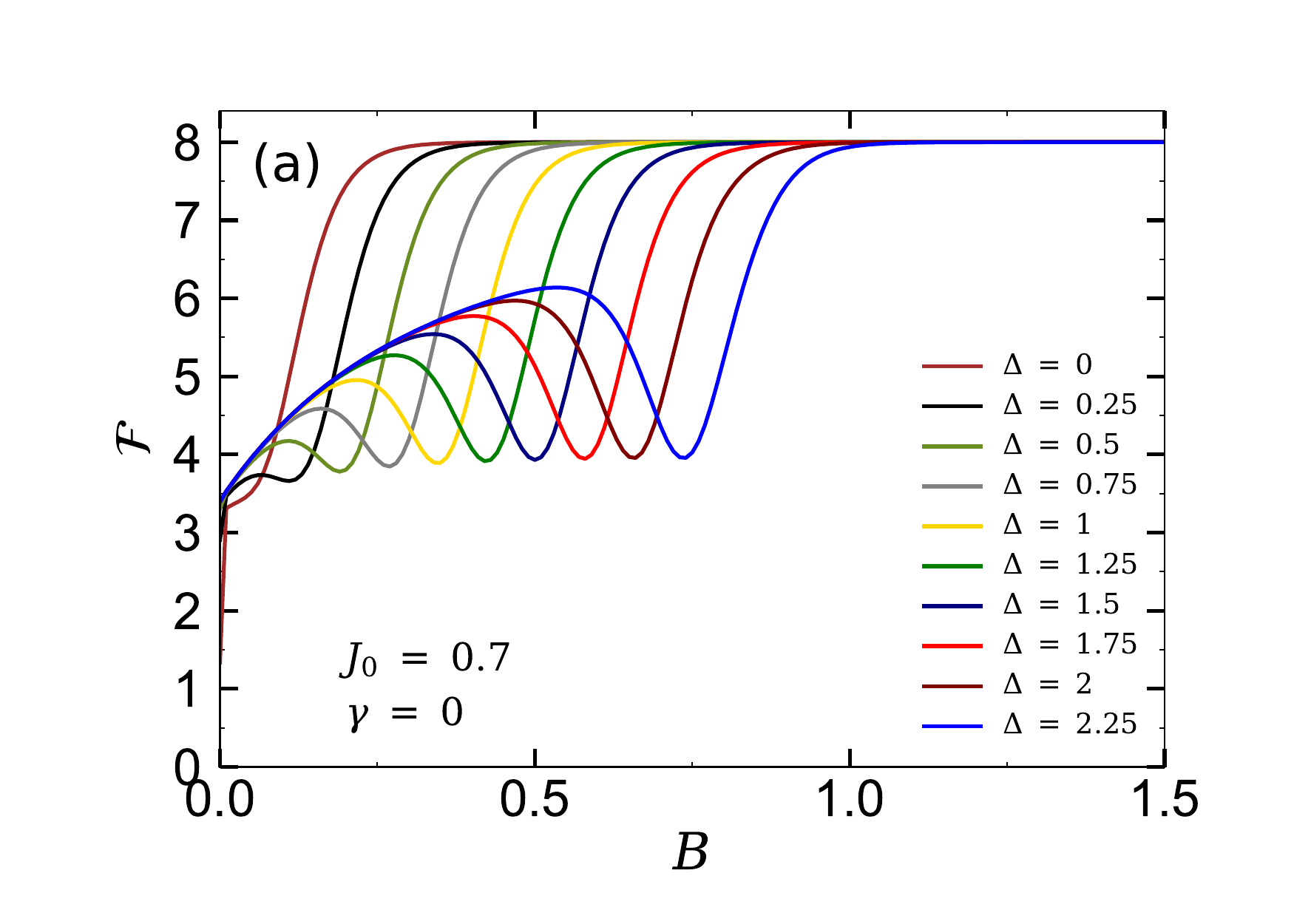}
\includegraphics[scale=0.45]{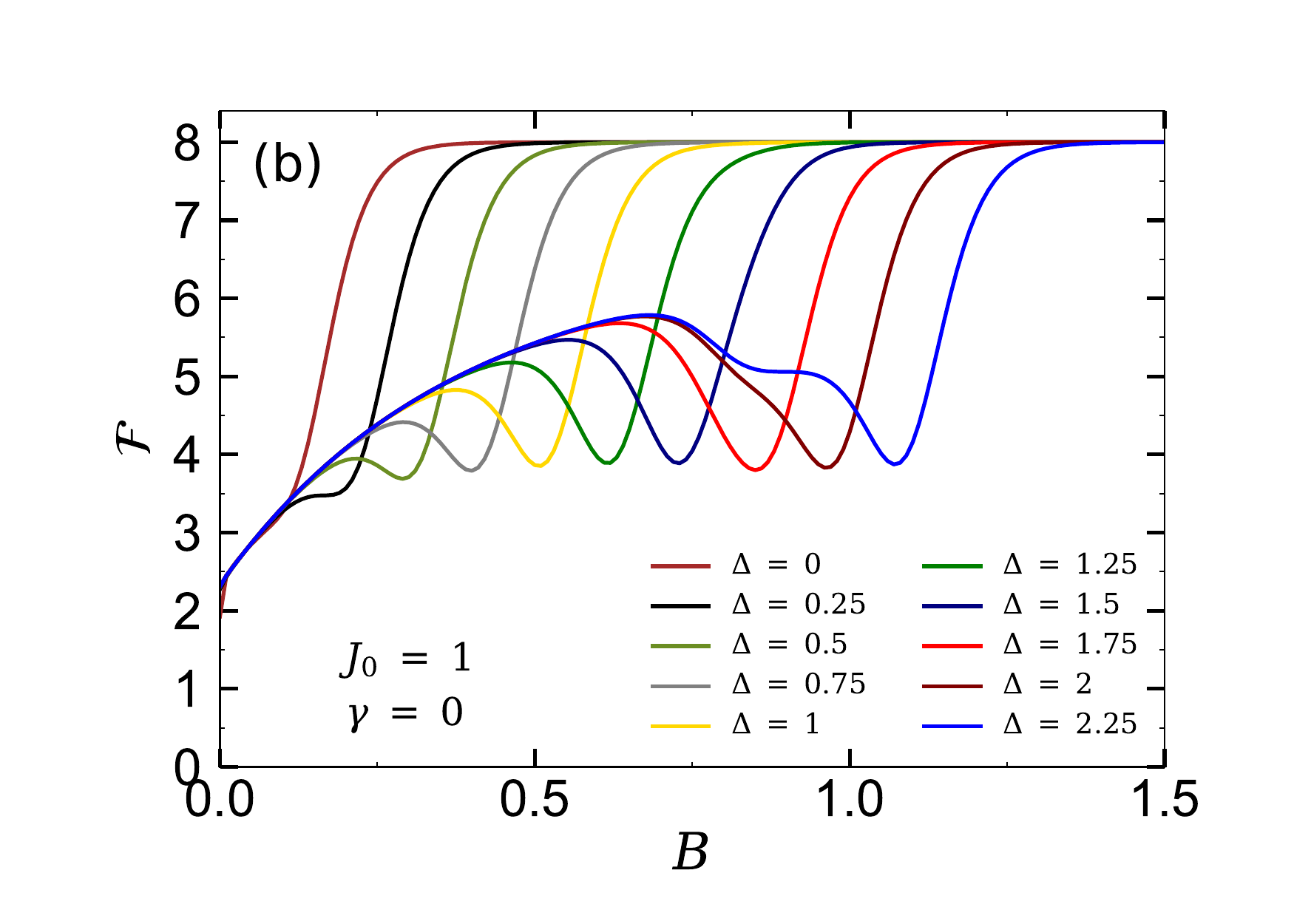}
\includegraphics[scale=0.45]{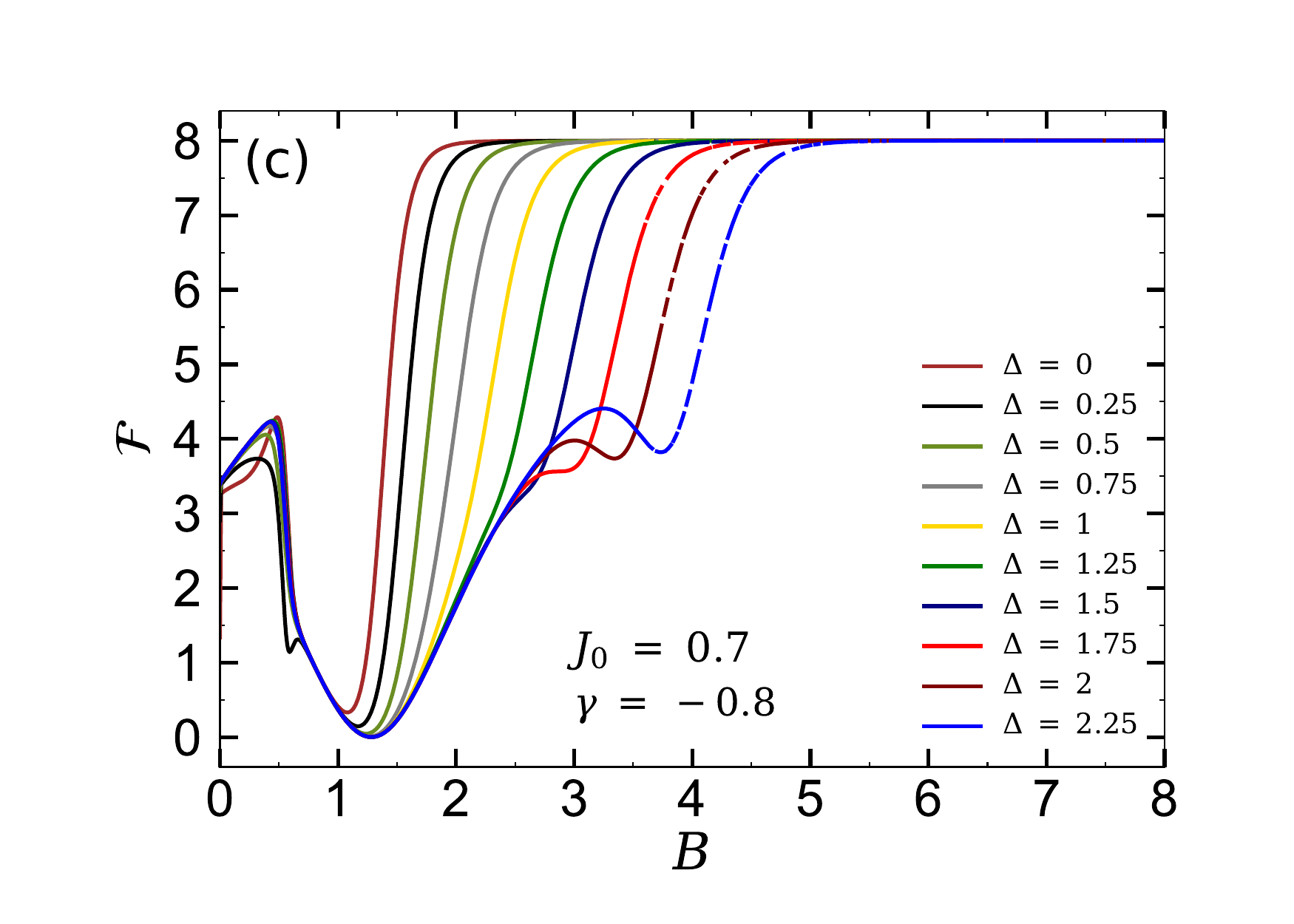}
\includegraphics[scale=0.45]{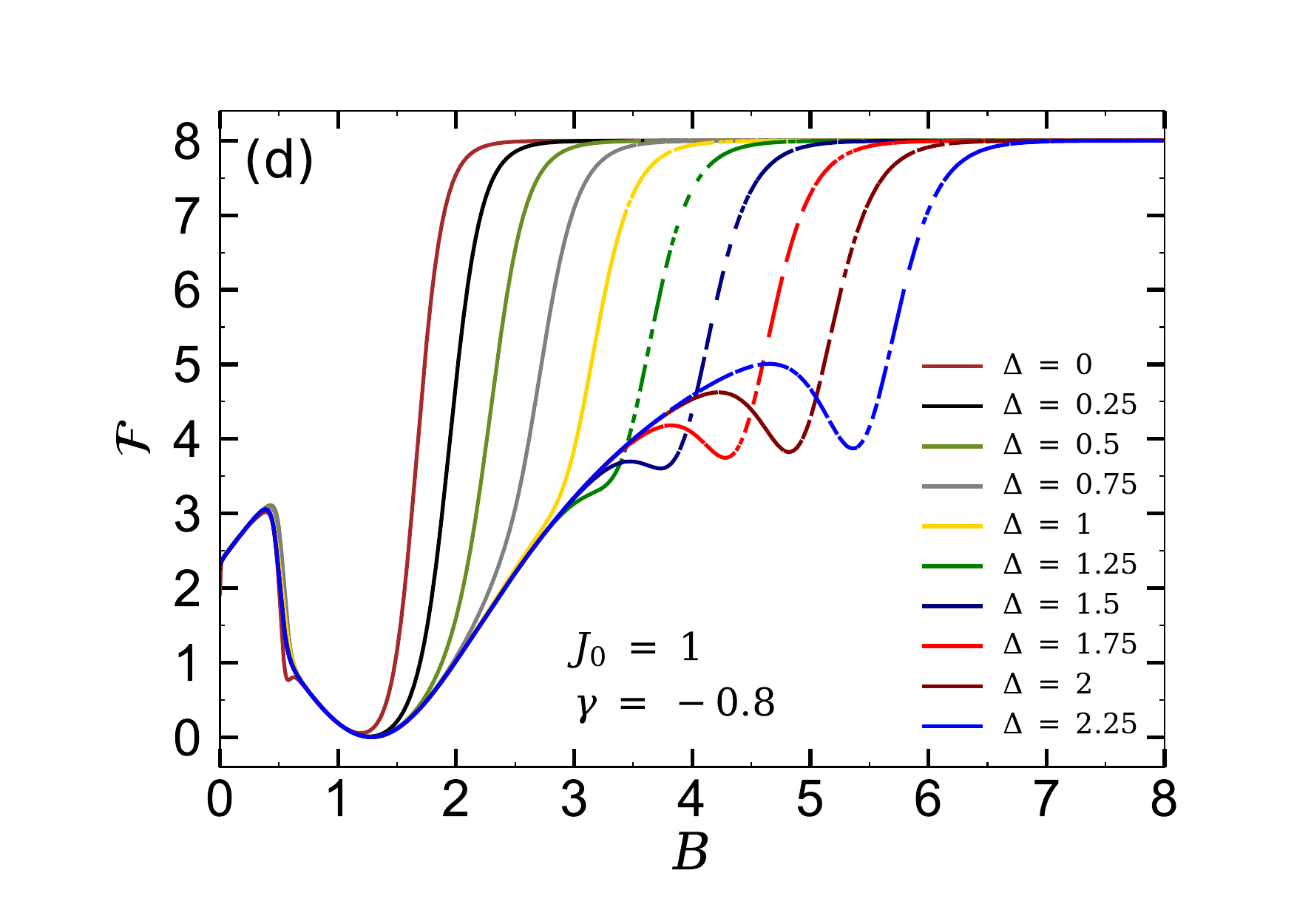}
\caption{\label{QFI} The QFI as a function of magnetic field $B$ for several fixed values of the anisotropic parameter $\Delta$.
(a) and (b) correspond to the original model in the absence of impurity ($\gamma=0$), assuming $J_0=0.7$ and  $J_0=1$, respectively.
 (c) and (d) show QFI  of the model with magnetic impurity, $\gamma=-0.8$, assuming $J_0=0.7$ and  $J_0=1$, respectively. 
 In all panels we considered low temperature $T=0.05$. We fixed other parameters as in Fig. \ref{Fig3}.
 We note that here and in the next figure, the discontinuity in lines related to the high exchange anisotropy $\Delta>1$ and high magnetic fields happens for the sake of we calculated the QFI numerically and does not result in losing the accuracy of the subject.}
\end{figure}

Figures \ref{QFI}(a) and \ref{QFI}(b) illustrate the QFI of the original model ($\gamma=0$) for various fixed values of $\Delta$ and two selected values of exchange interaction $J_0=0.7$ and $J_0=1$, respectively. It is observable that $\mathcal{F}$ increases monotonically upon increase of the magnetic field, then reaches an intermediate minimum. The magnetic position of this minimum depends on the exchange anisotropy $\Delta$. In other words, by increasing  $\Delta$, the mentioned minimum appears at higher magnetic fields.  For all of considered $\Delta$ there is a steep increase of the QFI in a special magnetic field interval. This function ultimately reaches its maximum value at sufficiently high magnetic fields. By inspecting Fig.  \ref{QFI}(b) one sees that as the Ising coupling constant $J_0$ increases, the magnetic position of the produced minimum shifts towards higher magnetic fields. Moreover, for high values of the $\Delta$ there are two intermediate minima (see blue line).

One of the most interesting and novel results for the Ising-Heisenberg spin chain of the heterotrimetallic coordination compound 
$\mathrm{Fe-Mn-Cu}$ with magnetic impurity is crystal clear in Figs.  \ref{QFI}(a) and \ref{QFI}(b). We see that the QFI behavior is significantly different from the original case. In fact, by inducing a magnetic impurity into the model, the QFI increases with increase of the magnetic field till reaches a typical maximum at very low magnetic fields. Then sharply drops down at critical magnetic field $B_c\approx 0.5$.
Surprisingly, the QFI reaches its minimum value $\mathcal{F}=0$ at the critical magnetic field $B_c=1.282$. We remind that at this critical magnetic field the concurrence becomes maximum (review Fig. \ref{Fig3}). With further increase of the magnetic field, QFI shows a steep increase, reaching its maximum value for the exchange anisotropy range $\Delta\lesssim 1$. For the range $\Delta> 1$, the QFI makes an intermediate minimum at high magnetic fields afterwards reaches its maximum. Comparing panel \ref{QFI}(b)  with \ref{QFI}(a) corroborates that increase of the Ising nodal exchange interaction $J_0$ entails the QFI reaches its maximum at remarkably higher magnetic fields. However, the magnetic positions of the sharp dropping down and reaching minimum value  $\mathcal{F}=0$ do not change by altering $J_0$.

\begin{figure}
\includegraphics[scale=0.45]{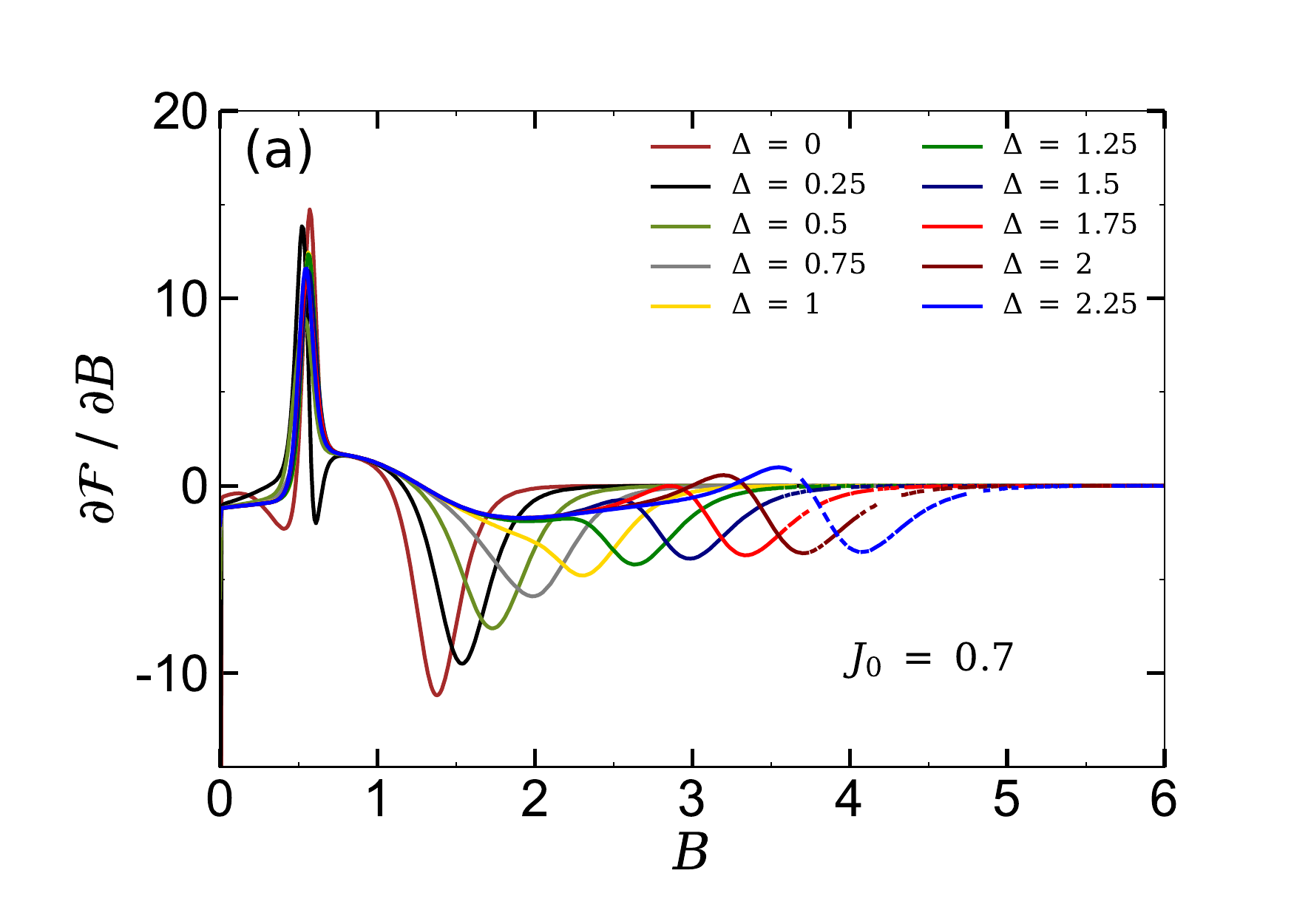}
\includegraphics[scale=0.45]{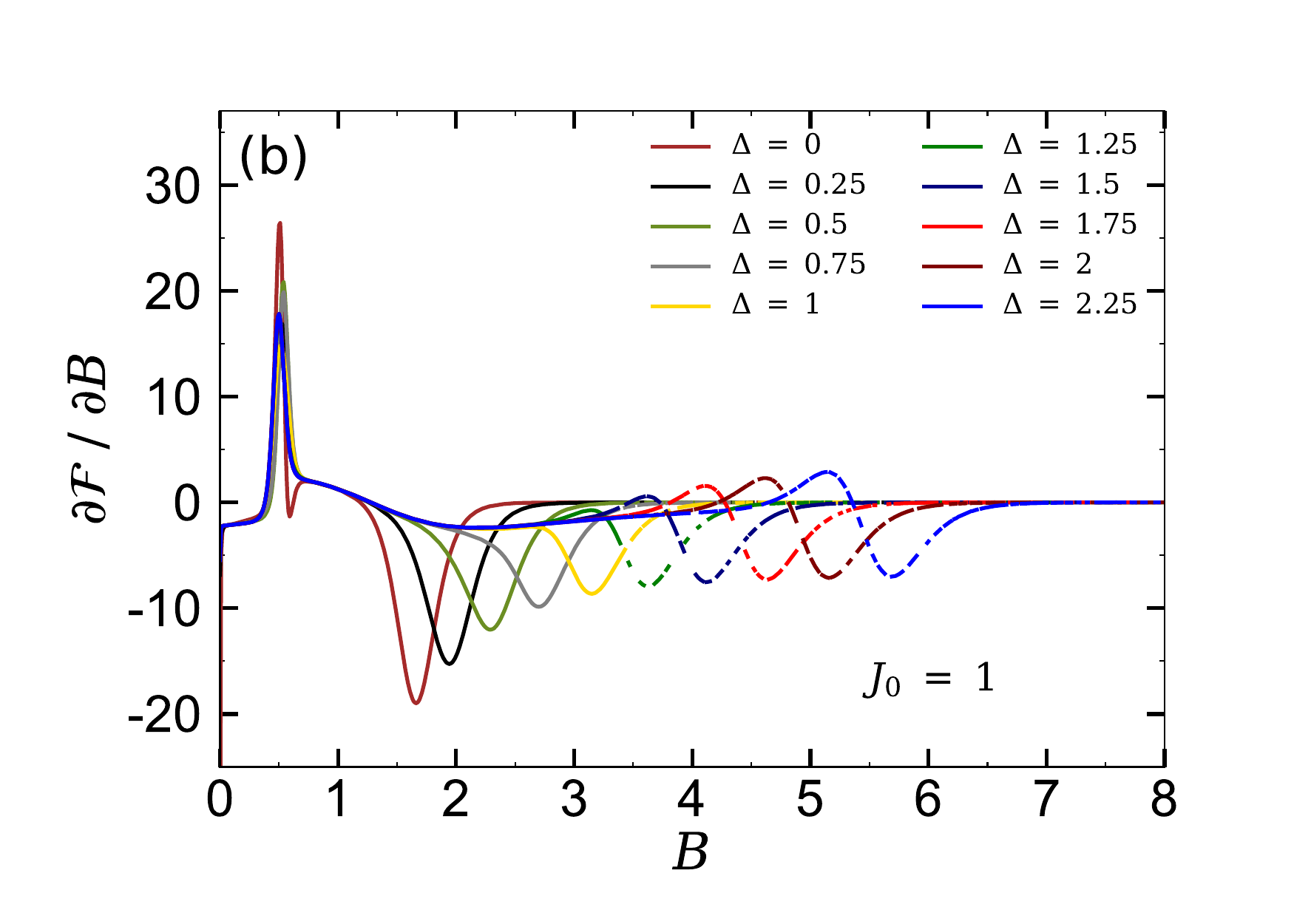}
\caption{\label{dBQFI} The first magnetic field derivative of the QFI of the model with impurity ($\gamma=-0.8$) against the magnetic field for several fixed values of $\Delta$. (a) $J_0=0.7$ and  (b) $J_0=1$.  In both panels we supposed low temperature $T=0.05$ for the model. Other parameters have been taken as in Fig. \ref{Fig3}.}
\end{figure}

The first magnetic field derivative of the QFI of the model with impurity as a function of the magnetic field has been displayed in Figs. \ref{dBQFI}(a) and \ref{dBQFI}(b), assuming fixed $J_0=0.7$ and $J_0=1$, respectively. The QFI shows a sharp pick at critical  magnetic field $B_c\approx 0.5$. According to the expressions in Ref. \cite{Ye2020}, one can observe that regardless of the value $J_0$ a quantum phase transition happens at $B_c\approx 0.5$.

\section{Quantum teleportation}\label{QT}

\begin{figure}
\includegraphics[scale=0.7,trim=10 120 0 80, clip]{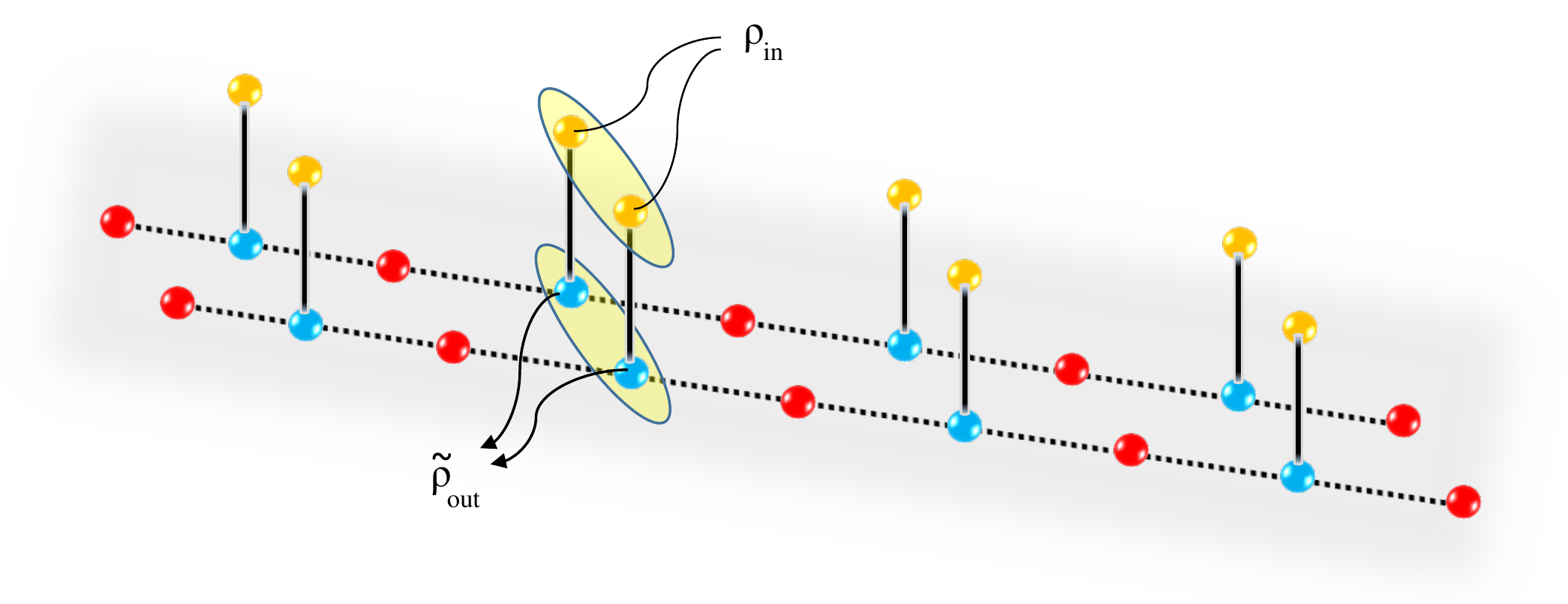}
\caption{\label{Fig7} The schematic representation  for the teleportation of the input state $\rho_{in}$ through a couple of independent quantum channels $(\rho_{ch})$. The teleported output state is denoted by $\widetilde{\rho}_{out}$.}
\end{figure}

In this section, we study the quantum teleportation by means of an entangled mixed state as resource. The standard teleportation can be regarded as a general depolarizing channel \cite{bow} and we investigate the influence of the magnetic impurity of the Ising-Heisenberg spin chain of the heterotrimetallic coordination compound $\mathrm{Fe-Mn-Cu}$ on the teleportation.
We suppose the input state that would be teleported has the form
\begin{equation}
|\psi_{in}\rangle=\cos\left(\frac{\theta}{2}\right)|10\rangle+e^{i\phi}\sin\left(\frac{\theta}{2}\right)|01\rangle\;,
\end{equation}
where $0\leq\theta\leq\pi$ and $0\leq\phi\leq2\pi$. Here $\theta$ describe all states with different amplitudes and $\phi$ are phase of these states, and in the density operator formalism, the state input is $\rho_{in}=|\psi_{in}\rangle\langle\psi_{in}|$. The concurrence $\mathcal{C}_{in}$ of the input state can be written as 
 \begin{equation}
\mathcal{C}_{in}=2\;\vert e^{i\phi}\sin\left(\frac{\theta}{2}\right)\cos\left(\frac{\theta}{2}\right)\vert=|\sin(\theta)|.
\end{equation}
When  a two-qubit state $\rho_{in}$ (as depicted in Fig. \ref{Fig7})  is teleported via the mixed channel $\widetilde{\rho}_{ch}$ of the two independent Ising-Heisenberg spin chain of the heterotrimetallic coordination compound $\mathrm{Fe-Mn-Cu}$, the output state $\widetilde{\rho}_{out}$ is given by 

\[
\widetilde{\rho}_{out}=\sum_{i,j=\left\{ 0,1,2,3\right\} }p_{i}p_{j}\left(\sigma_{i}\otimes\sigma_{j}\right)\rho_{in}\left(\sigma_{i}\otimes\sigma_{j}\right)\;,
\]
where $\sigma^0=I$, $\sigma^1=\sigma^x$, $\sigma^2=\sigma^y$ and $\sigma^3=\sigma^z$ are Pauli matrices.
$p_{i}=tr\left[E^{i}\widetilde{\rho}_{ch}\right]$, $E^{0}=|\Psi^{-}\rangle\langle\Psi^{-}|$,
$E^{1}=|\Phi^{-}\rangle\langle\Phi^{-}|$, $E^{2}=|\Phi^{+}\rangle\langle\Phi^{+}|$, $E^{3}=|\Psi^{+}\rangle\langle\Psi^{+}|$, for which $\left\{|\Psi^{\pm}\rangle, |\Phi^{\pm}\rangle\right\}$ denote the Bell states. Here, we consider the density operator channel as  $\widetilde{\rho}_{out}\equiv\widetilde{\rho}(T)$.

Therefore, we can write the output state as

\begin{equation}
\widetilde{\rho}_{out}=\left[\begin{array}{cccc}
c & 0 & 0 & 0\\
0 & f & \kappa & 0\\
0 & \kappa & g & 0\\
0 & 0 & 0 & c
\end{array}\right]\;,\label{eq:rho-out}
\end{equation}
where
\begin{flushleft}
\[
\begin{array}{cl}
c= & \left(\widetilde{\rho}_{2,2}+\widetilde{\rho}_{3,3}\right)\left(\widetilde{\rho}_{1,1}+\widetilde{\rho}_{4,4}\right),\\
f= & \left(\widetilde{\rho}_{1,1}+\widetilde{\rho}_{4,4}\right)^{2}\cos^{2}\left(\frac{\theta}{2}\right)+\left(\widetilde{\rho}_{2,2}+\widetilde{\rho}_{3,3}\right)^{2}\sin^{2}\left(\frac{\theta}{2}\right),\\
g= & \left(\widetilde{\rho}_{2,2}+\widetilde{\rho}_{3,3}\right)^{2}\cos^{2}\left(\frac{\theta}{2}\right)+\left(\widetilde{\rho}_{1,1}+\widetilde{\rho}_{4,4}\right)^{2}\sin^{2}\left(\frac{\theta}{2}\right),\\
\kappa= & 2e^{i\phi}\widetilde{\rho}_{2,3}^{\,2}\sin\theta.
\end{array}
\]
\par\end{flushleft}

To get a better understanding of the quantum teleportation and average fidelity, let us first describe the thermal entanglement of the output state $\mathcal{C}_{out}$. Using Eq. (\ref{eq:rho-out}) in the definition of concurrence (\ref{eq:conco}), we obtain the output concurrence $\mathcal{C}_{out}$ defined by
\[
\mathcal{C}_{out}=2\max\left\{ 2\widetilde{\rho}_{2,3}^{\,2}\mathcal{C}_{in}-|\widetilde{\rho}_{2,2}+\widetilde{\rho}_{3,3}||\widetilde{\rho}_{1,1}+\widetilde{\rho}_{4,4}|,0\right\} .
\]
In the following section, we analyze the quality of the process of the teleportation.

\subsection{Average fidelity of teleportation}\label{AFT}

In this section, we turn our attention to the quality of the process of teleportation. The fidelity between $\rho_{in}$ and $\rho_{out}$ characterizes the quality of the teleported state. The fidelity is defined by \cite{joz}
\[
F=\langle\psi_{in}|\rho_{out}|\psi_{in}\rangle\;.
\]
After some straightforward algebra, the fidelity can be described as
\begin{equation}
F=\frac{\sin^{2}\theta}{2}\left[\left(\widetilde{\rho}_{1,1}+\widetilde{\rho}_{4,4}\right)^{2}+4\widetilde{\rho}_{2,3}^{\,2}-\left(\widetilde{\rho}_{2,2}+\widetilde{\rho}_{3,3}\right)^{\,2}\right]+\left(\widetilde{\rho}_{2,2}+\widetilde{\rho}_{3,3}\right)^{\,2}.\label{eq:fidelity}
\end{equation}
When the input state is a pure state, the efficiency of quantum communication is characterized by average fidelity. We can calculate the average fidelity as

\[
F_{A}=\frac{1}{4\pi}\intop_{0}^{2\pi}d\phi\intop_{0}^{\pi}F\sin\theta d\theta\;.
\]

\begin{figure}

\includegraphics[scale=0.40]{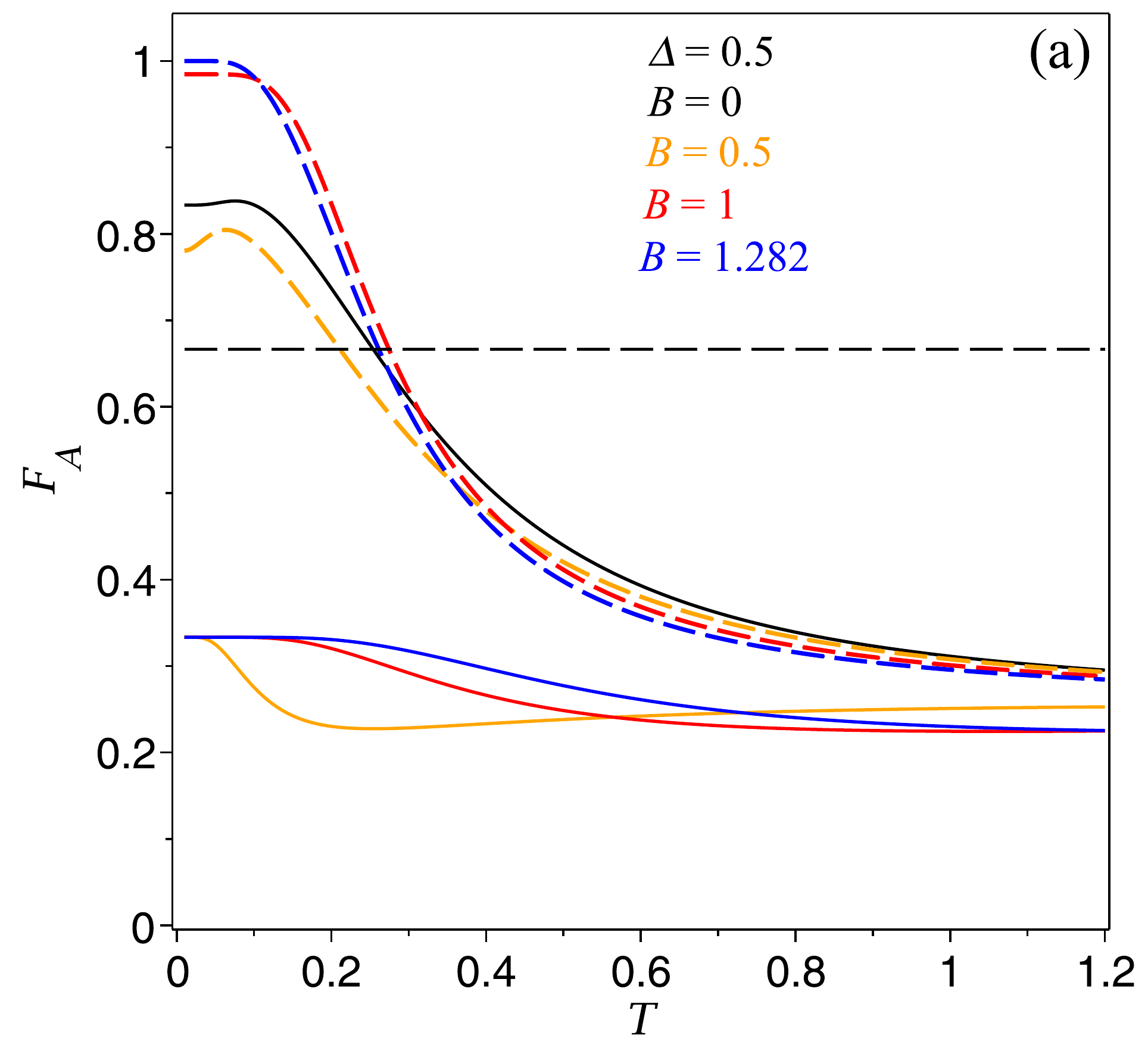}
\includegraphics[scale=0.40]{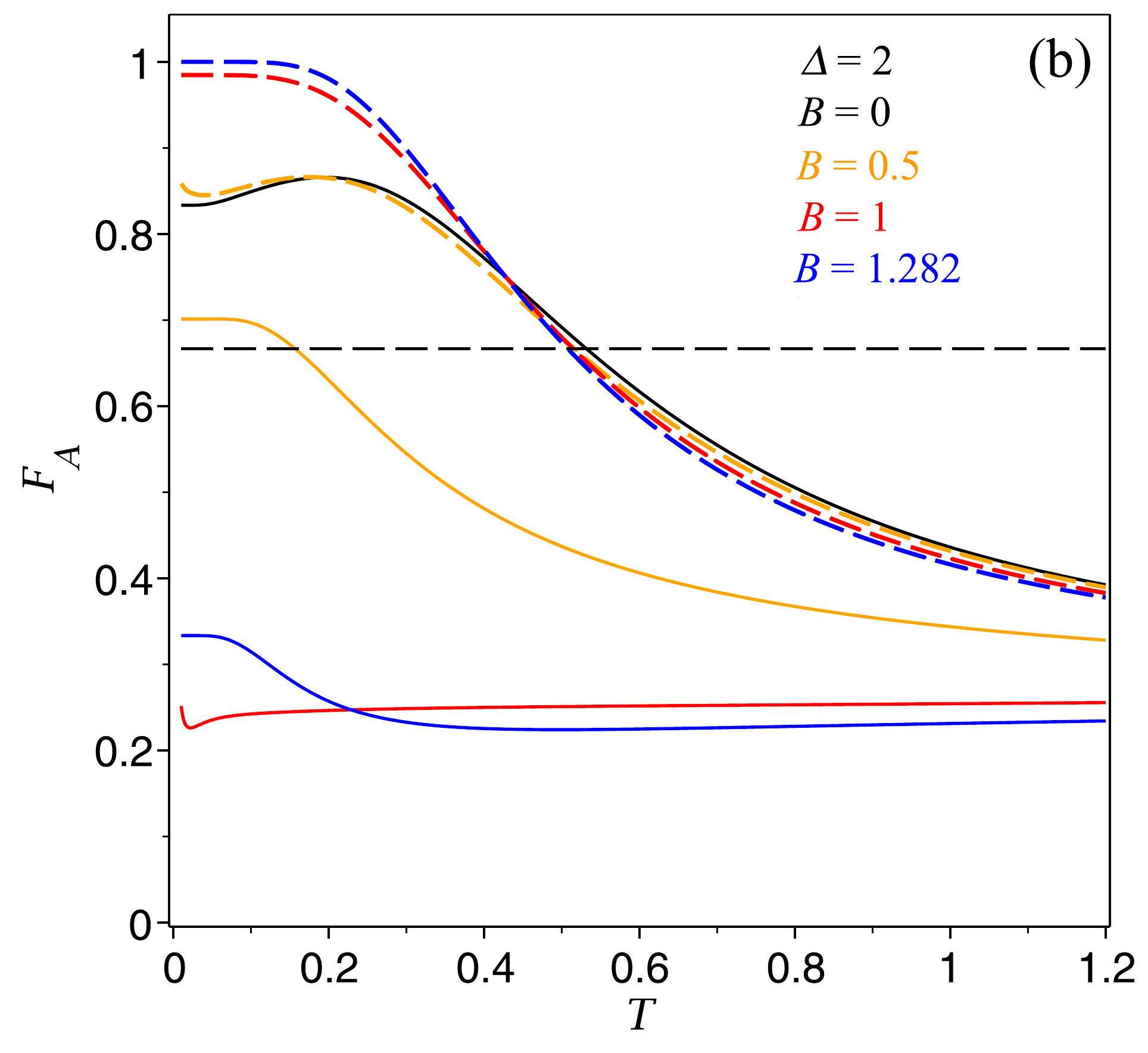}
\caption{\label{Fig8} The average fidelity $F_{A}$ as a function of temperature $T$ for several fixed values of the magnetic field and parameter set (a) $\Delta=0.5$, (b) $\Delta=1$, assuming fixed $J_0=1$, $J=1$ and impurity parameter $\gamma=-0.8$.}

\end{figure}

For the model described here, the average fidelity $F_{A}$ can be given by
\begin{equation}
F_{A}=\frac{1}{3}\left[\left(\widetilde{\rho}_{1,1}+\widetilde{\rho}_{4,4}\right)^{2}+4\widetilde{\rho}_{2,3}^{\,2}-\left(\widetilde{\rho}_{2,2}+\widetilde{\rho}_{3,3}\right)^{\,2}\right]+\left(\widetilde{\rho}_{2,2}+\widetilde{\rho}_{3,3}\right)^{\,2}.\label{eq:fidelityaverage}
\end{equation}
To transmit the input state better than any classical communication protocol, $F_{A}$ must be greater than $\frac{2}{3}$  which is the best fidelity in the classical world.
In Fig. \ref{Fig8}(a)  we show the behavior of average fidelity $F_{A}$ in terms of temperature $T$ for several fixed values of the magnetic field, assuming $J_{0}=1$ and $\Delta=0.5$ and impurity parameter $\gamma=-0.8$. The horizontal dashed lines at $F_{A}=2/3$ denote the limit of quantum fidelities. In Fig. \ref{Fig8}(a), we can see that for this choice of parameters the only possibility of teleportation of information happens for null magnetic field in the original model. Meanwhile, when we insert magnetic impurity, we have a considerable improvement in quantum teleportation. 
It is quite noteworthy that for the critical point $B_{max}=1.282$ the average fidelity becomes maximal at sufficiently low temperatures. It can be seen that average fidelity is above 2/3 for our model with magnetic impurity in contrast to the original model where it is not possible to teleport information.

In Fig. \ref{Fig8}(b) is  depicted the temperature dependence of the average fidelity for $\Delta=2$. For this choice of anisotropy parameter, the original model allows teleportation for weak magnetic fields and lower temperatures. For example, for $B=0.5$ (see orange solid curve), as soon as the temperature increases, the average fidelity falls down below $2/3$, signaling the impossibility of teleportation. On the other hand, the fidelity behavior is significantly different for the case when a magnetic impurity is embedded. 
Our results show major improvement in the average fidelity that can be achieved by tuning the strength of the magnetic impurity.
In conclusion, we can take up the way of inducing magnetic impurities as a creative technique to manipulate and to enhance the teleportation processing.

\begin{figure}

\includegraphics[scale=0.40]{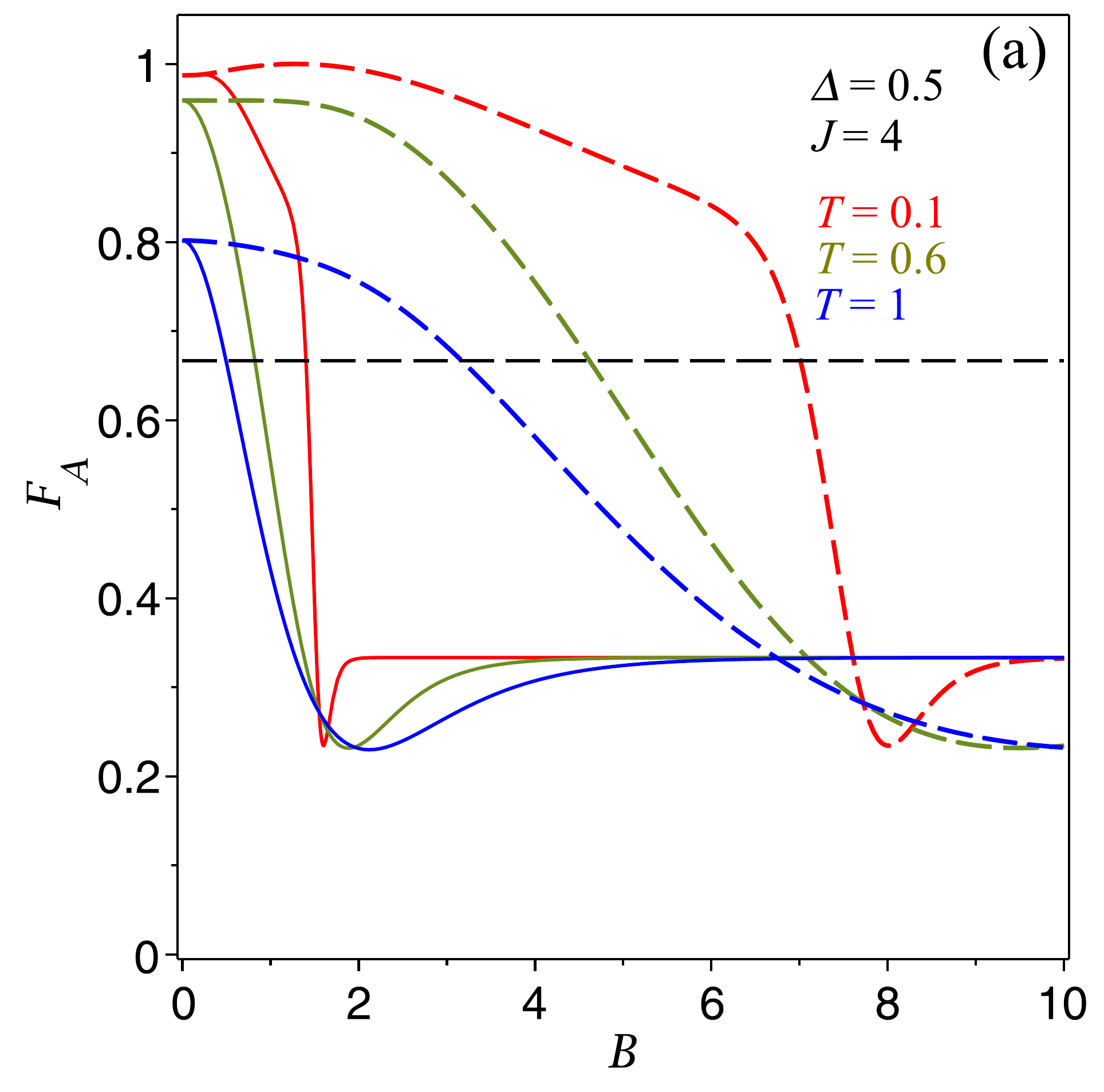}
\includegraphics[scale=0.40]{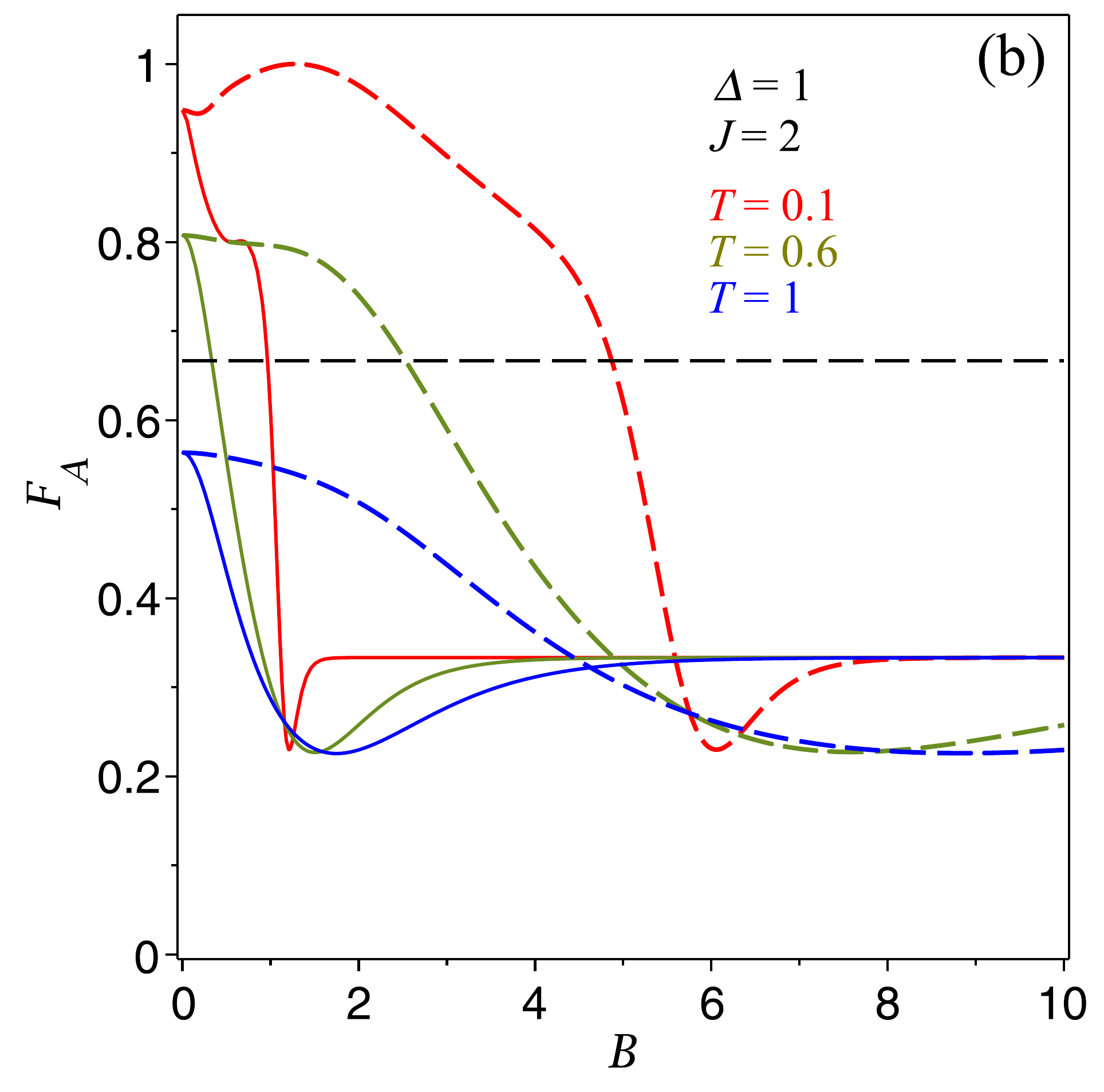}
\caption{\label{Fig10} The average fidelity $F_{A}$ as a function of magnetic field $B$ for several fixed values of the temperature and parameter set (a) $J_0=1$, $J=4$, $\Delta=0.5$, and (b) $J_0=1$,  $J=2$, $\Delta=1$, assuming impurity parameter $\gamma=-0.8$.}

\end{figure}

Finally, in Fig. \ref{Fig10}, the average fidelity $F_{A}$ is plotted as a function of magnetic field $B$ for three selected temperatures $T=0.1$, $T=0.6$ and $T=1$ and $\gamma=-0.8$. In panels of this figure, solid lines describe the average fidelity of the original model recently investigated by Y. -D. Zheng {\it et al.} \cite{zheng}.
 In Fig. \ref{Fig10}(a), we fixed $J_0=1$, $J=4$ and $\Delta=0.5$. 
As can be seen, the average fidelity remains above $2/3$ for weak magnetic fields in the original model \cite{zheng}. When the impurity is induced in the model, in contrast to the original case, the behavior of the average fidelity becomes more robust, enabling teleportation of information in the regions of very strong magnetic fields. 
%Under this condition, one can immediately have two stimulating observations when the original model is considered.
%First, the average fidelity sharply decreases upon increasing the magnetic field. Secondly, the teleportation of information is only possible for weak magnetic fields. However, by assuming the inclusion of the magnetic impurity (dashed curves), we have a dramatic enhancement of the average fidelity $F_{A}$. In particular, for the low-temperature regime (i.e., $T=0.1$),  the inclusion of magnetic impurity generates an increase in the average fidelity until reaching maximum fidelity ($F_{A}=1$). Then, this function decreases monotonically when the magnetic field increases further. However, for the higher temperature ($T=1$), the average fidelity remains below $2/3$, making it impossible the existence of the quantum teleportation  of information. 

Analogously,  Fig. \ref{Fig10}(b) illustrates the average fidelity $F_{A}$ for the case $J_0=1$, $J=2$ and $\Delta=1$.
Under this condition, one can immediately have two stimulating observations when the original model is considered.
First, the average fidelity sharply decreases upon increasing the magnetic field. Secondly, the teleportation of information is only possible for weak magnetic fields. However, by assuming the inclusion of the magnetic impurity (dashed curves), we have a dramatic enhancement of the average fidelity $F_{A}$. In particular, for the low-temperature regime (i.e., $T=0.1$),  the inclusion of magnetic impurity generates an increase in the average fidelity until reaching maximum fidelity ($F_{A}=1$). Then, this function decreases monotonically as the magnetic field increases further. However, for the higher temperature ($T=1$), the average fidelity remains below $2/3$, making it impossible the existence of the quantum teleportation  of information. 
All of these findings show again that a considerable enhancement of the teleportation of information can be achieved by tuning the strength of the magnetic impurity for the Ising-Heisenberg spin chain of the heterotrimetallic coordination compound $\mathrm{Fe-Mn-Cu}$. 

\section{Conclusions} \label{Conclusions}

In this work we have examined the influences of a typical magnetic impurity on the quantum properties of the Ising-Heisenberg spin-1/2 chain of the heterotrimetallic coordination compound $\mathrm{Fe-Mn-Cu}$. 
At first step, we have considered a magnetic impurity on a local
Heisenberg dimer of the chain. Next, we have exactly solved  the model within the transfer-matrix formalism and paid our attention  to the study the thermal pairwise entanglement and $l_1$-norm of the coherence as measures of the quantum correlation of the impurity dimer.
 We witnessed a large enhancement on the both of thermal entanglement and the quantum coherence when the model involves with a magnetic impurity.
One of our notable results is that the thermal pairwise entanglement can be controlled and tuned by imposing a magnetic impurity into the model.

 By verifying the QFI of the model we understood that the inclusion of the magnetic impurity substantially affects on the behavior of this function. Furthermore, QFI presented anomalous behavior nearby the critical magnetic fields $B_c\approx 0.5$ and $B_c=1.282$. Close to the first critical point, QFI showed a sharp value, but at second one it vanished. Rely on the investigations carried out in this work, we proved that
QFI can be considered as a useful quantum tool for estimating the quantum phase transition of the Ising-Heisenberg spin-1/2 chain of the heterotrimetallic coordination compound $\mathrm{Fe-Mn-Cu}$. 

 The teleportation scenario for the two-qubits in a typical quantum state has been also discussed through a couple of
quantum channel including a Heisenberg dimer with magnetic impurity. Each of channels has been assumed to be constructed by an infinite Ising-Heisenberg spin-1/2 chain of heterotrimetallic coordination compound $\mathrm{Fe-Mn-Cu}$. 
%We observed  that the average fidelity of teleportation is significantly enhanced when the model possesses magnetic impurity.  
We have demonstrated that at low-temperature regime, the inclusion of magnetic impurity leads to remarkable enhancement in the average fidelity until reaching maximum value.
Based on our findings regarding  the model with magnetic impurity,  we claimed that the average fidelity becomes more robust compared with the original model, enabling teleportation of information in the regions of very strong magnetic fields.
%We saw that it is possible to teleport information in a wide range of anisotropic models. This is impossible in the model without any impurity. As a final word, we state that considerable enhancement of the teleportation can be achieved by tuning the strength of the impurity parameters. This can be used locally to control the quantum resources and the quantum teleportation of the information, unlike the original model where it is globally done.
%Our notable result is that a considerable enhancement of the teleportation of information can be achieved by tuning the strength of the magnetic impurity for the Ising-XXZ spin chain of the heterotrimetallic coordination compound $\mathrm{Fe-Mn-Cu}$.  
\section{Acknowledgments}

H. Arian Zad acknowledges the receipt of the grant from the Abdus Salam International Centre for Theoretical Physics (ICTP), Trieste, Italy, and the CS MES RA in the frame of the research project No. SCS18T-1C155. 
This work was partially supported by CNPq, CAPES and Fapemig. M. Rojas would like to thank CNPq grant 432878/2018-1.
The authors are also grateful to Prof. N. Ananikian for his insightful discussion.

%\section*{References}

\end{document}